\documentclass[preprint,12pt]{aastex}

\shorttitle{HD 141569 Association}
\shortauthors{Aarnio et al.}
\usepackage{lscape}

\begin{document}

\title{A SURVEY FOR A COEVAL, COMOVING GROUP ASSOCIATED WITH HD 141569}

\author{Alicia N.\ Aarnio\altaffilmark{1}, Alycia J.\ Weinberger\altaffilmark{2}, Keivan G.\ Stassun\altaffilmark{1}, Eric E. Mamajek\altaffilmark{3}, and David J. James\altaffilmark{1,4}}

\altaffiltext{1}{Department of Physics \& Astronomy, Vanderbilt University, Nashville, TN 37235}
\altaffiltext{2}{Department of Terrestrial Magnetism, Carnegie Institution of Washington, 5241 Broad Branch Road NW, Washington, DC 20015}
\altaffiltext{3}{Harvard-Smithsonian Center for Astrophysics, 60 Garden Street, Cambridge, MA 02140}
\altaffiltext{4}{Department of Physics \& Astronomy, University of Hawaii at Hilo, 200 W. Kawili Street, Hilo, HI 96720}

\begin{abstract}
We present results of a search for a young stellar moving group associated 
with the star HD 141569, a nearby, isolated Herbig AeBe primary member of a 5$\pm$3 Myr-old 
triple star system on the outskirts of the Sco-Cen complex.  Our spectroscopic survey 
identified a population of 21 Li-rich, $\lesssim$30 Myr-old stars within 30\degr\ of 
HD 141569 which possess similar proper motions with the star.  The spatial distribution 
of these Li-rich stars, however, is not suggestive of a moving group associated with the 
HD 141569 triplet, but rather this sample appears cospatial with Upper Scorpius and 
Upper Centaurus Lupus.  We apply a modified moving cluster parallax method to 
compare the kinematics of these youthful stars with Upper Scorpius and Upper Centaurus 
Lupus.  Eight new potential members of Upper Scorpius and five new potential members of 
Upper Centaurus Lupus are identified.  A substantial moving group with an identifiable 
nucleus within 15\degr ($\sim$30 pc) of HD 141569 is not found in this sample.
Evidently, the HD 141569 system formed $\sim$5 Myr ago in relative isolation, tens 
of parsecs away from the recent sites of star formation in the Ophiucus-Scorpius-Centaurus region.

\end{abstract}


\section{INTRODUCTION}\label{intro}
The HD 141569 stellar system initially garnered the interest of 
\citet{Rossiter:1943} as a potential triple star system.  The 
existence of this group was confirmed over half a century later when 
\citet{Weinberger:2000} showed that the stars' position angles relative 
to one another had not changed since Rossiter's 1938 observations, thus 
confirming common proper motions.  They also noted that all three stars 
are consistent with being the same age, indicating a comoving and coeval 
system.  This triumvirate is particularly interesting because 
all components are quite young; the system was found to be 5$\pm$3 Myr 
\citep{Weinberger:2000}, and recently the A star's age was 
constrained using surface gravities and effective temperatures
to be $\sim$4.7 Myr \citep{Merin:2004}.
HD 141569 itself is near enough to resolve its large, dusty disk, and 
near-infrared signatures indicate perturbations to the disk's 
structure which could be explained by planet formation \citep{Weinberger:1999}.
Were a coherent group present, it would therefore be distinctly possible 
to similarly observe additional young, disk-bearing stars.  Furthermore, 
surveying a coeval sample at this distance would be useful for determining 
disk frequency and mechanisms by which they arise and evolve.

In some cases, seemingly isolated young stars have later been found to 
have an entourage of other low-mass young stars which constitute a stellar 
association \citep[e.g. HD 104237, HR 4796, $\beta$ Pictoris; cf.][]{Mamajek:2002,Li:2005}.  
The lower mass members of these associations have been useful, indeed critical, for 
estimating the age of the massive star in question, as often the massive 
star is on the main sequence and uncertainty in its HR diagram position is 
substantial enough to impede an isochronal age estimate.  Any additional 
association members should be identifiable by their similar space motions.  The youth 
of these objects makes them useful for studies of circumstellar disks and 
stellar evolution.  In this region of sky rich with 
already identified associations such as $\rho$ Ophiucus, Upper Scorpius (US) and Upper 
Centaurus Lupus (UCL), placing an ``HD 141569 Association'' into context with surrounding 
moving groups could aid in the understanding of star formation histories in giant 
molecular clouds.

So motivated, in this paper we seek to identify a sample of stars associated 
with HD 141569.  Previous studies \citep[e.g.][]{Mamajek:2002,Li:2005}
have shown that catalog searches based on X-ray activity, proper motions, and 
distance criteria can yield new members of associations. Here we also identify 
candidate members with a catalog search for X-ray active stars in the vicinity of 
HD 141569 possessing proper motions consistent with HD 141569 (\S\ref{targets}). 
From new spectroscopic observations (\S\ref{obs}), in \S\ref{analysis} we identify a 
subset of 21 stars which we claim as youthful using Li {\sc I} equivalent widths; 
furthermore, we also note the presence of H$\alpha$ emission as an interesting quantity 
usually indicative of chromospheric activity. Lacking direct distance measurements for 
these stars, we derive distances using a modified version of the traditional 
moving cluster parallax method \citep{deBruijne:1999, deZeeuw:1999, Mamajek:2005},
with which we further determine the probability that the sample stars are 
comoving with HD 141569 (\S\ref{mcp}).

Placing these stars on a Hertzsprung-Russell diagram (hereafter HRD or HR diagram, \S\ref{sec_hrd}),
we find that the stars' distances---and hence ages---are most consistent with 
their strong lithium abundances when we derive their moving cluster parallaxes assuming 
comovement with Upper Scorpius or UCL associations of young stars.  
In \S\ref{summconc} we summarize our study and present thirteen newly identified Sco-Cen 
members.

\section{DATA}\label{data}

\subsection{Target Selection}\label{targets}
\begin{figure}[ht]
\plotone{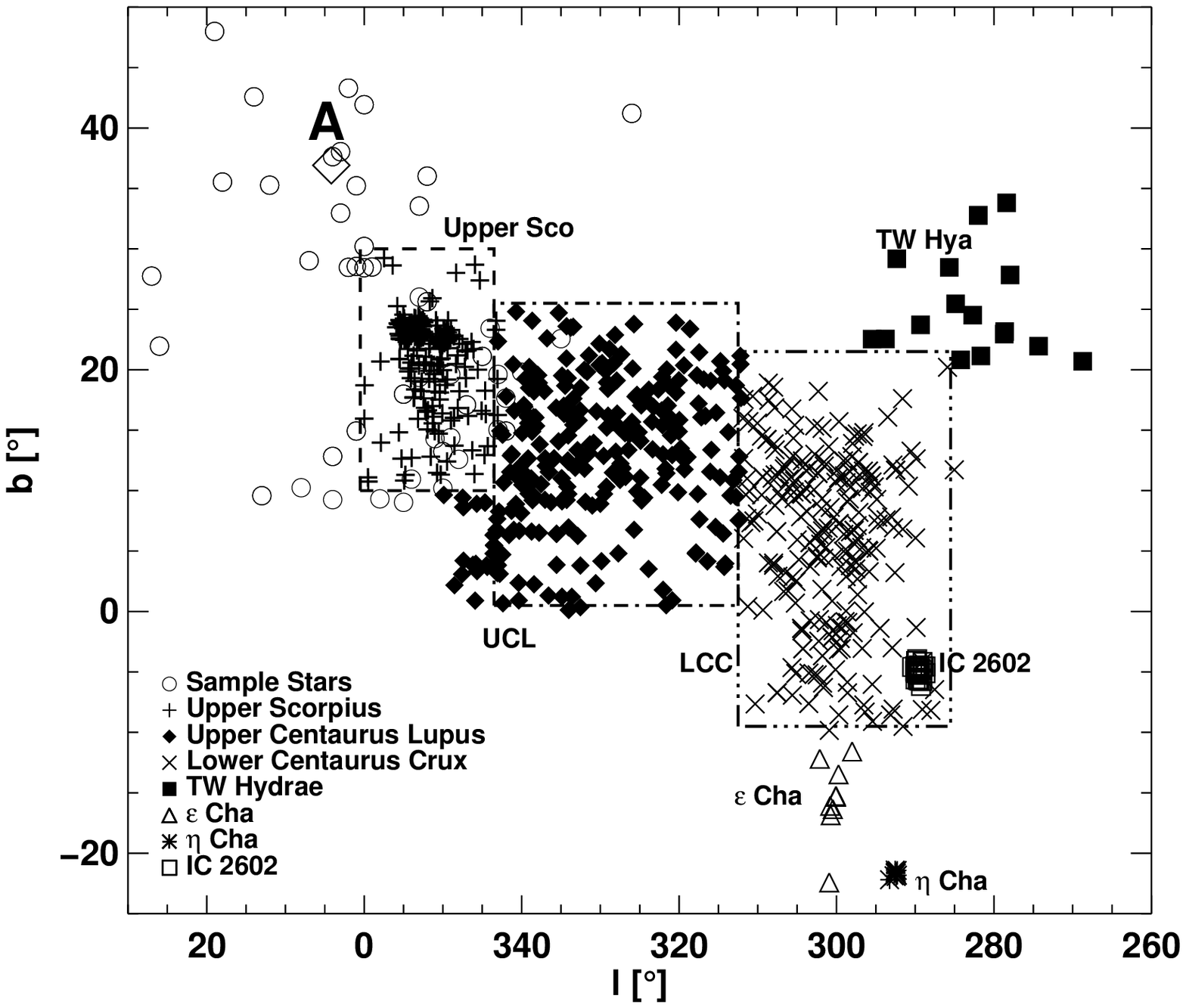}
\caption{\label{fig_galcoords}
A galactic coordinate map of HD 141569 and nearby associations is presented.  Open circles 
indicate our sample stars, while dashed boxes indicate regions studied by \citet{deZeeuw:1999}. 
Details of objects plotted herein are to be found in the following papers: 
Upper Scorpius \citep{Preibisch:2002}, UCL and LCC \citep{Mamajek:2002,deZeeuw:1999}, TW Hya 
\citep{Mamajek:2005}, $\eta$ and $\epsilon$ Cha \citep{Zuckerman:2004,Zuckerman:2001} and IC 2602
\citep{Robichon:1999}.  In all subsequent plots, the letter A denotes the position 
of HD 141569.}
\end{figure}

Potential targets for observation were found by two catalog queries for stellar properties indicative 
of youth and comovement with HD 141569.  In the first search, 10 stars were found through a query of the 
\textit{Hipparcos} catalog \citep{Hipparcos:1997} for objects within 10\degr\ of HD 141569 with similar 
distances \citep[99$\pm$8pc, as reported in][]{Hipparcos:1997} and proper motions.  Eight objects from the 
{\sc \textit{ROSAT}} Faint Source Catalog \citep{Voges:2000} supplemented this sample, as previous 
low-resolution observations (A. Weinberger, unpublished data) showed evidence for spectroscopic signatures 
indicative of youth.  A second search was performed in which the {\sc \textit{ROSAT}} Bright Source 
Catalog \citep{Voges:1999} was probed for X-ray sources within a 30\degr\ radius of HD 141569.  Of the 
1,114 resulting targets, $\sim$400 sources had Tycho-2 \citep{Hog:2000} catalogued proper motions.  We 
required proper motions to be within
$\pm 15$ and $\pm 20$ mas yr$^{-1}$ in RA and Dec of HD 141569's proper motion, respectively,
based on the range of proper motions observed in the widely dispersed and
similarly aged TW Hydrae Association \citep{Zuckerman:2001,Webb:1999}.  Finally, we required 
that the \textit{ROSAT} sources not be extended and that they be cospatial with 
the Tycho coordinates.  This procedure resulted in $\sim$70 objects desirable 
for further study to establish youth and space motions.  A map of the observed sample 
is shown in Fig.\ \ref{fig_galcoords}; members of other nearby associations are also 
shown for added context and perspective.

\subsection{Observations}\label{obs}
Forty-nine (49) stars of our input catalog were observed spectroscopically over the course 
of three observing runs in order to measure spectral types, radial velocities, and Li {\sc I} and 
H$\alpha$ equivalent widths.  Four targets were found to be close visual 
binaries so their companions were observed as well; of these four, 
two were determined to be background objects on the basis of apparent 
magnitudes and spectral type in comparison to the primary
and are thus not discussed further.  An observing log of the 51 stars 
(49 targets $+$ 2 visual companions) which form our final sample for this work
is presented in Table \ref{obslog}. Photometry, proper motions, and 
parallax measurements from the literature are documented in Table \ref{StellarParams}.

On 2001 June 18 (UT), the ten \textit{Hipparcos} selected targets and five FSC targets 
were observed using the Hamilton Echelle Spectrometer at Lick Observatory.  
The Hamilton echelle covers a wavelength range of $\sim$3500-10000\AA\ 
with a resultant resolving power of 60,000 at 6000\AA\ when using a slit width of 
1.2\arcsec.  During April 2002, a further 36 target stars 
of the second catalog search were observed using the echelle spectrograph 
on the Ir\'{e}n\'{e}e du Pont telescope at Las Campanas Observatory.  The du Pont 
spectrograph observes a wavelength range of $\sim$3700-9800\AA\ with a resolving power 
of 40,000 at a slit width of 0.75\arcsec.  On both observing runs, comparison arc 
spectra were taken to establish a pixel to wavelength calibration.  At LCO, Thorium-Argon 
lamp spectra were taken between each object exposure because the spectrograph is situated 
at the Cassegrain focus, whereas the Coud\'{e}-fed Hamilton echelle required less 
frequent arc calibration, with arc spectra be taken at the beginning and end of the night.

\subsection{Reduction Procedure}
The echelle data were reduced using {\sc IRAF}\footnote{Image Reduction and 
Analysis Facility, IRAF, is distributed by the National Optical Astronomy Observatories, 
which are operated by the Association of Universities for Research in Astronomy, Inc., under 
cooperative agreement with the National Science Foundation.} to perform 
standard spectral reduction procedures.  Instrumental effects were accounted 
for via measurement of bias level and read noise, variations in pixel-to-pixel 
{\sc CCD} response were removed in the flat fielding process, and dead columns were 
identified (which then prevented measurement of the H$\alpha$ 6563 \AA\ line).  
A basic reduction process was then followed to locate and extract the echelle 
orders as well as remove scattered light.  The Thorium-Argon arcs were similarly 
extracted, their features identified so as to wavelength calibrate the object 
spectra.  Examples of {\sc IRAF} routines employed include \textit{apall}, 
\textit{ecidentify}, \textit{refspec}, and \textit{dispcor}.

\section{ANALYSIS}\label{analysis}

We aim to identify the subset of our sample which shows evidence for youth consistent 
with membership to a young, $\sim$5 Myr-old association.  Additionally, we wish to assess 
kinematic properties and test for comovement.  To these ends, from the literature, we 
obtain effective temperatures as well as near infrared photometry and proper motions.  
From the spectra, we measure Li {\sc I} equivalent widths as well as radial velocities.  
Utilizing these data, we estimate ages and select a youthful, high lithium sample, and we 
test for kinematic similarity against the velocity models of nearby, young moving groups.

\subsection{Effective Temperatures}\label{teff}

Spectral types for the majority of our sample are reported in the literature 
\citep{Houk:1982,Houk:1988,Houk:1999,Torres:2006}. These are converted to 
$T_{\rm eff}$ using the main-sequence spectral type--$T_{\rm eff}$ relationship of 
\citet{Kenyon:1995}. We supplement these with spectral types determined from 
low-resolution spectra obtained with the Lick KAST spectrograph (A.\ Weinberger, 
unpublished data) as well as $T_{\rm eff}$ determined from the \ion{Fe}{1} to 
\ion{Sc}{1} line ratio \citep[cf.][]{Stassun:2004a,Steffen:2001} observed in our 
high-resolution spectra (\S\ref{obs}). The line ratios measured for our spectral 
type standards are in good agreement with the calibration of \citet{Basri:1990}; 
we thus adopt their line ratio-spectral type scale in assigning types to our sample.  
In some cases, primarily for more massive stars, the \ion{Fe}{1} and \ion{Sc}{1} 
lines were not present or could not be measured with confidence above the noise 
level.  $T_{\rm eff}$ values are summarized in Table \ref{Spectroscopy}.  
Where multiple $T_{\rm eff}$ values are available, they generally show good consistency 
with one another to within $\sim 300$ K (corresponding to $\sim 2$ spectral subclasses). 
For our final set of effective temperatures, we adopt literature spectral types where 
available, line ratio spectral types if not.  For one object, HD 157310B, neither 
were available, and thus we interpolated its 2MASS \citep[2 Micron All Sky Survey,][]{Skrutskie:2006} $(H-K_{s})$ color over the effective 
temperature-color relationship of \citet{Kenyon:1995}.  The set of adopted temperatures 
is reported in the final column of Table \ref{Spectroscopy}.

\begin{figure}[ht]
\plotone{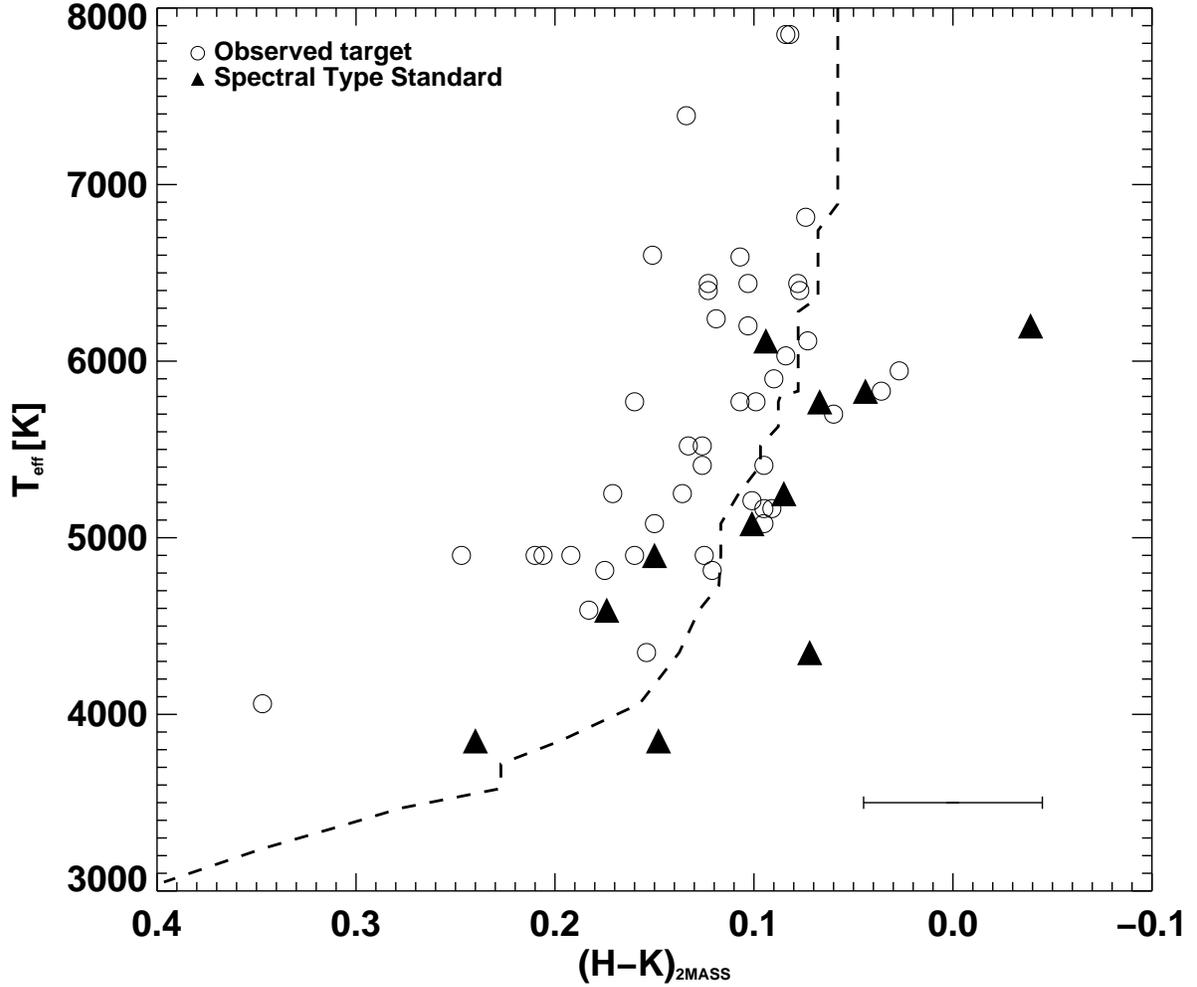}
\caption{\label{fig_tempscale} 
Color-$T_{\rm eff}$ relationship from \citet{Kenyon:1995} for 2MASS $(H-K)$ colors.
The solid line represents the main-sequence relationship as defined in \citet{Kenyon:1995}.  
Open circles represent literature spectral types for our sample as reported in column 6 of 
Table \ref{Spectroscopy}.  Filled triangles represent literature spectral types for the standard 
stars observed by us (Table \ref{obslog}).
}
\end{figure}

\begin{figure}[ht]
\plotone{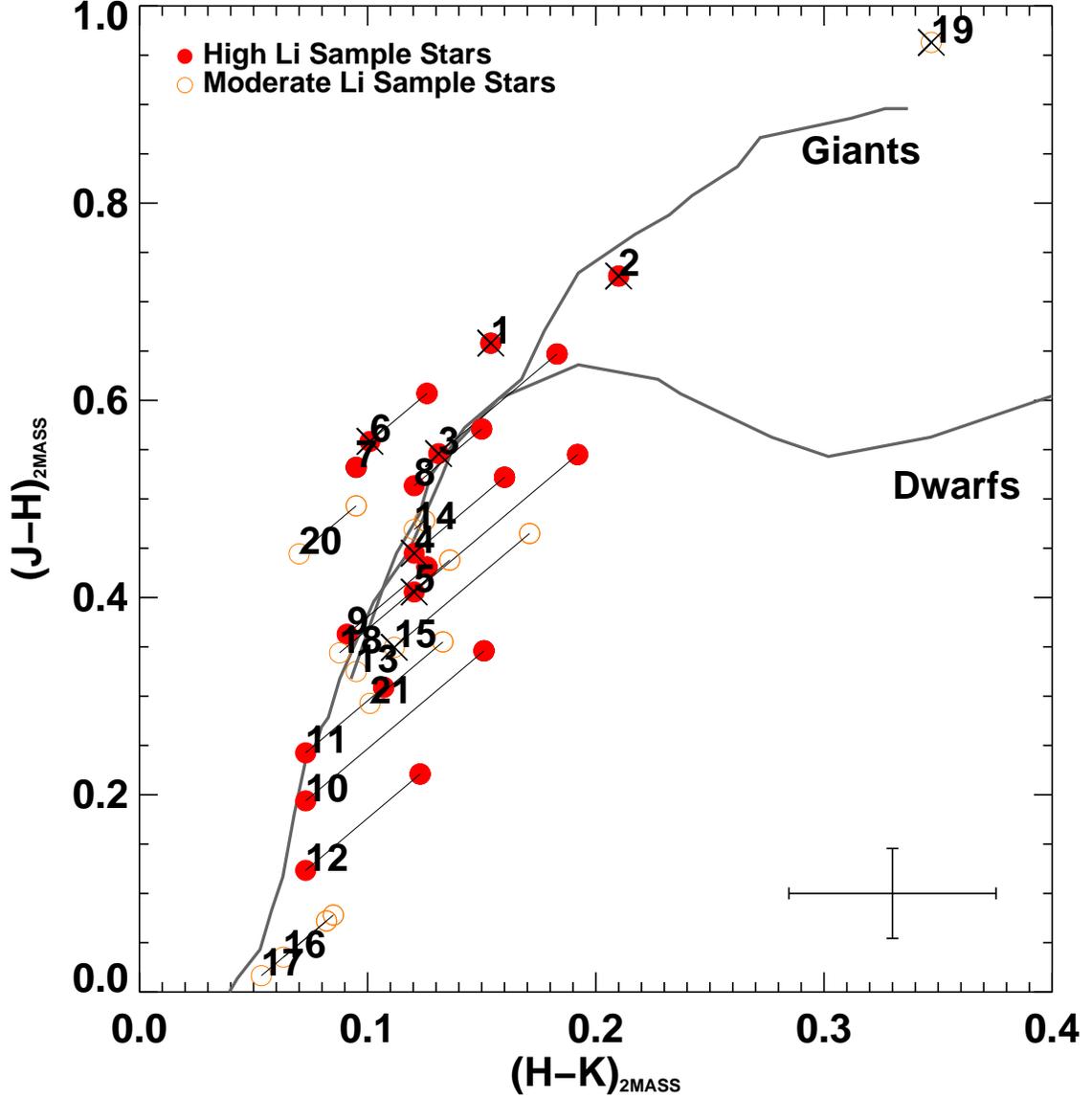}
\caption{\label{fig_JHHK}
Near-infrared $JHK$ colors from 2MASS. For visual clarity, we display here only the 21 stars 
that we identify as being lithium rich (see \S\ref{lithium} and Fig.\ \ref{fig_lithium}).
Solid lines represent the dwarf and giant sequences from \citet{Bessell:1991} and \citet{Bessell:1988}.  
Dereddened objects are plotted with observed and dereddened colors connected by a thin, solid line 
parallel to a reddening vector defined by $\frac{E_{(J-H)}}{E_{(H-K)}}=1.95$ \citep{Bessell:1988}. 
Numbers next to the points are provided for ease in identifying the objects in the data tables.  
Stars with H$\alpha$ in emission are indicated with an X over the point.}

\end{figure}

In Fig.\ \ref{fig_tempscale} we show these effective temperatures as a function of 
the objects' observed 2MASS $(H-K_{s})$ colors; also plotted is the $T_{\rm eff}- (H-K_{s})$ 
relationship from \citet{Kenyon:1995}. For comparison, the standard stars observed 
(Table \ref{obslog}) are also shown in this parameter space. The observed $(H-K_{s})$ colors 
follow the expected relationship with $T_{\rm eff}$ with a scatter of $\sim 300$ K,
consistent with the scatter in $T_{\rm eff}$ from spectral types above. This
indicates that our sample in general suffers relatively little extinction. 
Indeed, radio survey column-density measurements \citep{Kalberla:2005} in the direction 
of our targets indicate that $E(H-K_{s})$ reddening toward our sample should be $\lesssim 0.1$ mag.  
Via comparison with expected intrinsic $(H-K_{s})$ for a star of given effective temperature, 
we derive $K_{s}$ band extinction values, $A_{Ks}$, and deredden the sample accordingly.  
$(H-K_{s})$ colors are plotted in Fig.\ \ref{fig_JHHK}; lines illustrate dereddening by 
connecting colors before and after dereddening.  Also displayed for comparison are 
the dwarf sequence from \citet{Bessell:1988} as well as reddening vectors assuming
the standard ratio of total-to-selective extinction $R_V = 3.12$. For visual clarity,
we only display the 21 targets identifed as lithium rich (\S\ref{lithium}).
The $K_{s}$-band extinction corresponding to the applied dereddenings 
is in all cases $A_{Ks} < 0.15$ mag, with the exceptions of 2MASS J17215666-2010498 and 
TYC 6191-0552-1 (objects \#19 and \#2 in the data tables) for which $A_{Ks}=0.33$ mag 
and $A_{Ks}=0.16$ mag, respectively.  We have checked \textit{Spitzer} 24$\mu$m data 
(A. Weinberger, private communication) and find that while most of the high Lithium sample 
lacks 24$\mu$m excess, 2MASS J17215666-2010498 has a substantial excess.  TYC 6191-0552-1 
was analyzed by \citet{Meyer:2008} and was found to have moderate 24$\mu$m excess. 
For both objects, apparent excess in the $H$ and $K_{S}$ bands in concert with 24um excess 
confirms disk presence and thus we cannot and do not apply a standard interstellar dereddening law.

\subsection{Lithium Equivalent Width}\label{lithium}

\begin{figure}[ht]
\plotone{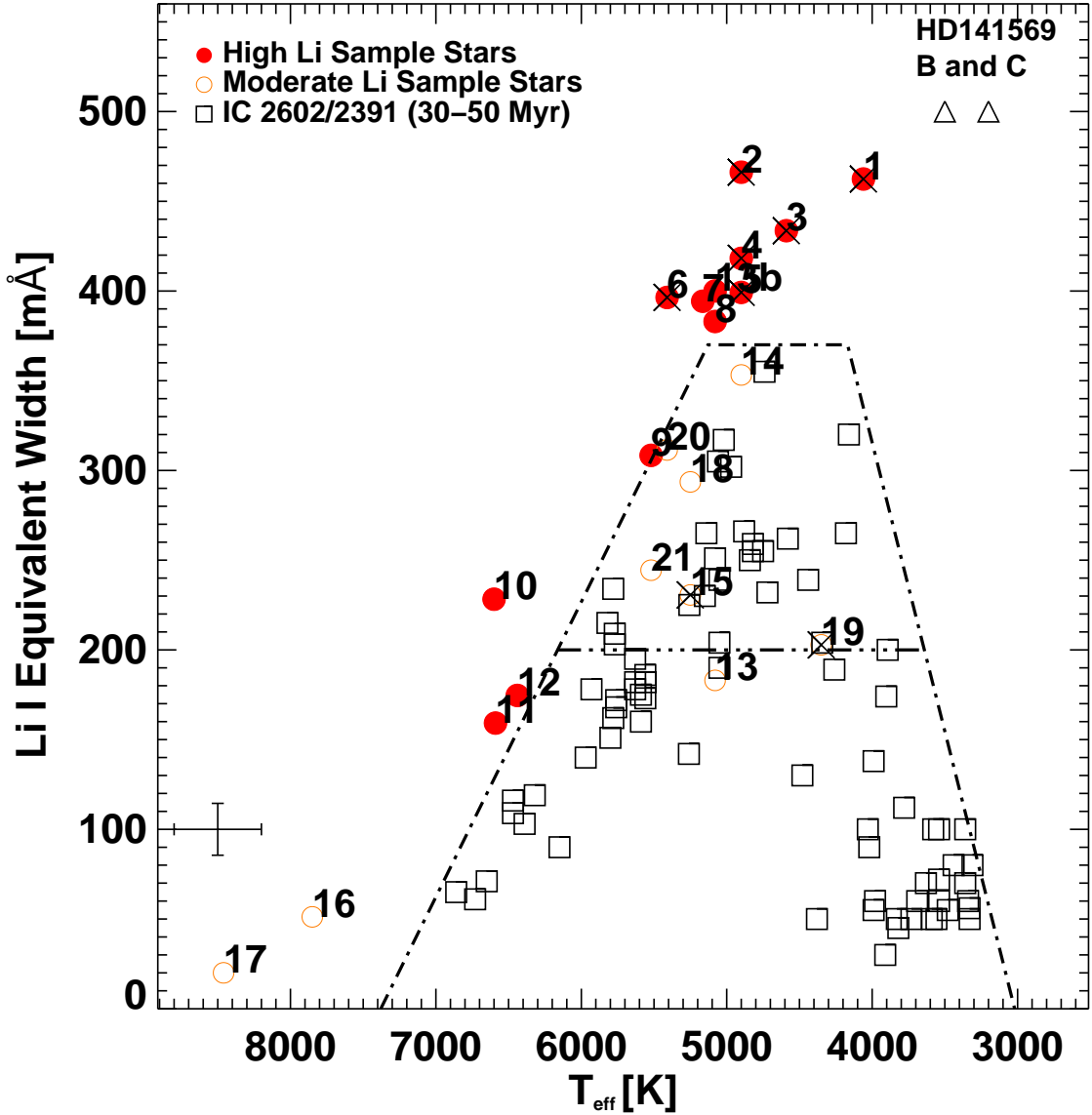}
\caption{\label{fig_lithium}
Fe contamination corrected Li {\sc I} EW as a function of T$_{\rm eff}$, plotted with the 
IC 2602/IC 2391 data from \citet{Randich:1997,Randich:2001}.  Stars possessing enough lithium 
to be above the selection threshold are likely to be pre--main-sequence, and are designated the
``High Li'' sample.  Stars below the threshold but above the horizontal line at 200m\AA\ 
constitute the ``Moderate Li'' sample; these stars have Li EWs in the upper envelope of 
the IC2602/IC2391 locus.  We include objects 16 and 17 in the moderate sample to be conservative 
as Li {\sc I} is not a good age indicator in higher mass stars.  Numbers next to the points are 
provided for ease in identifying the objects in the data tables.  In two cases we show both the 
measured EW as well as the doubled value (indicated by a ``b'' after the number identifier), 
these objects may be suffering Li line filling (see $\S$3.2).  Stars showing H$\alpha$ in emission 
are indicated with an X over the point.}
\end{figure}

We measured the equivalent width (EW) of the $\lambda$6707 line of Li {\sc I}
from our spectra using the IRAF routine \textit{Splot}. For each star, EW measurements 
were obtained by both directly integrating the flux in the line and by calculating
the area of a best-fitting Gaussian.  Our measured EW values include
contributions from the small \ion{Fe}{1}+CN line at 6707.44\AA, leading
to measured Li {\sc I} EWs that are representative of a slightly (10-20
m\AA) over-estimated photospheric Li presence. For instance \citet{Soderblom:1993}
report that this Fe line blend has an EW = [20(B-V)$_{0}$-3]m\AA\ for main sequence 
solar-type stars.  We correct for contamination following this prescription and find 
the median value for the sample is 10 m\AA; we report in Table \ref{Spectroscopy} 
the Li {\sc I} EWs for each target via both measurement methods as well as the 
Fe line blend contribution.  For all targets, the {\sc rms} of
the difference between the EWs determined using both methods is 18.4
m\r{A}. With the aim of conservatively selecting a sample, we utilize the lower of the 
two measured values; these are plotted in Fig. \ref{fig_lithium}.

To identify the young stars in the sample, Li {\sc I} EWs were compared to the 
upper envelope of EWs as a function of $T_{\rm eff}$ reported in the literature 
for the $\sim 30$ Myr-old clusters IC 2602 and IC 2391
\citep[][our Fig.\ \ref{fig_lithium}]{Randich:1997,Randich:2001}.
Twelve stars are found to have Li {\sc I} EWs above this envelope, compelling evidence of 
youth. In what follows we refer to these 12 stars as the ``High Li'' sample.

Also of interest, the H$\alpha$ profiles of several objects are in emission in our spectra 
and some possess double-peaked profiles. Another 6 stars show elevated lithium levels 
($\gtrsim 200$m\AA\ but are below the threshold shown in Fig.\ \ref{fig_lithium}), placing 
them in the upper envelope of the IC 2602 and IC 2391 loci.  An additional two stars have 
temperatures greater than 7500K and equivalent widths above the locus; although these are 
potentially older stars which simply lack deep enough convective zones to deplete primordial 
lithium abundances, we include them in the analyses for completeness.  We also include star 
\#13 in this group as it is in a double system separated by $\sim$1.2\arcsec while the seeing 
that night was $\sim$1.5\arcsec, making Li line filling likely.  In Fig. \ref{fig_lithium}, 
this star is plotted with its measured equivalent width doubled (denoted ``13b'') to 
demonstrate which sample group it would potentially belong to.  In what follows, we 
refer to these 9 stars as the ``Moderate Li'' sample, two of which show double-peaked H$\alpha$ 
in emission (Table \ref{Spectroscopy}).

These 21 stars, which are the most likely in our sample to be of comparable age to 
HD 141569, will be the focus of the remainder of our analyses. 
For ease in tracking these stars through our analysis, they are labeled with a
running numerical identifier in the data tables and figures, and all EW and line profile 
information is reported in Table \ref{Spectroscopy}.

\subsection{Radial Velocity}\label{rv}

Heliocentric radial velocities were obtained using the {\sc IRAF} task \textit{fxcor} to 
cross correlate each target spectrum against the radial-velocity standard star of 
closest spectral type (Table \ref{obslog}). Four echelle orders spanning the 
wavelength ranges 6025-6150\AA, 6150-6275\AA, 6625-6750\AA, and 5120-5220\AA\
were employed as they contain many deep metallic lines and little or no telluric contamination.
In Table \ref{Astrometry} we report the mean radial velocities from the four orders.

To determine the extent to which our radial-velocity measurements may be affected by 
the $v\sin i$ and signal-to-noise (S/N) of our target spectra, we performed a Monte 
Carlo simulation in which a narrow-lined, high S/N standard star spectrum was
randomly degraded one hundred times.  This process created artificially noisy spectra at 
S/N levels of 10--100, well representing the full range of S/N found in our sample.  
Each degraded spectrum was artificially broadened and cross-correlated against its 
original high S/N spectrum.  These degraded spectra were furthermore cross-correlated 
against the other standard stars to assess the effects of spectral-type mismatch on the 
resulting radial velocities.
We find that these effects are negligible (i.e.\ affecting the resulting
radial velocities by $\lesssim 1$ km s$^{-1}$) unless $v\sin i > 70$ km s$^{-1}$ or S/N $< 30$.
As all of our target spectra have S/N $>30$, only very fast rotators are potentially
affected (by up to 2.3 km s$^{-1}$ for $v\sin i = 100$ km s$^{-1}$).
Measured $v\sin i$ and radial velocity values are documented in Table \ref{Astrometry}.
In addition to these effects, we note that on an aperture-to-aperture basis, errors 
in wavelength calibration could affect the measured radial velocity; this can be 
quantified as the standard deviation of the mean radial velocity measured from the 
four selected apertures.  The final radial-velocity uncertainties quoted in Table 
\ref{Astrometry} are the quadrature sum of the internal uncertainty (the standard deviation 
of the four spectral orders used) and the uncertainty arising from rotational broadening. 

\subsection{Moving Cluster Parallaxes}\label{mcp}

With observed proper motions and measured radial velocities for the 21 ``High'' and 
``Moderate'' lithium stars, a kinematic picture of the sample is almost complete.  Tycho-2 
proper motions were used for consistency throughout (Table \ref{StellarParams}), the 
only exception being one of the two close visual binary companions, HD 157310B, for 
which only UCAC2 \citep{ucac2:2004} proper motions were available.  Proper motion data 
are not available for the other close companion, 2MASS J17215666-2010498; in what follows, 
we assume common proper motion with its primary star, TYC 6242-0104-1.  
\textit{Hipparcos} parallaxes are unavailable for the high lithium stars, thus 
a moving cluster parallax method \citep{deBruijne:1999} provides a means for 
determining their parallaxes and hence their distances.  With distances, it can be tested 
then whether these objects are consistent with being members of a coherent moving group.

\begin{figure}[ht]
\plotone{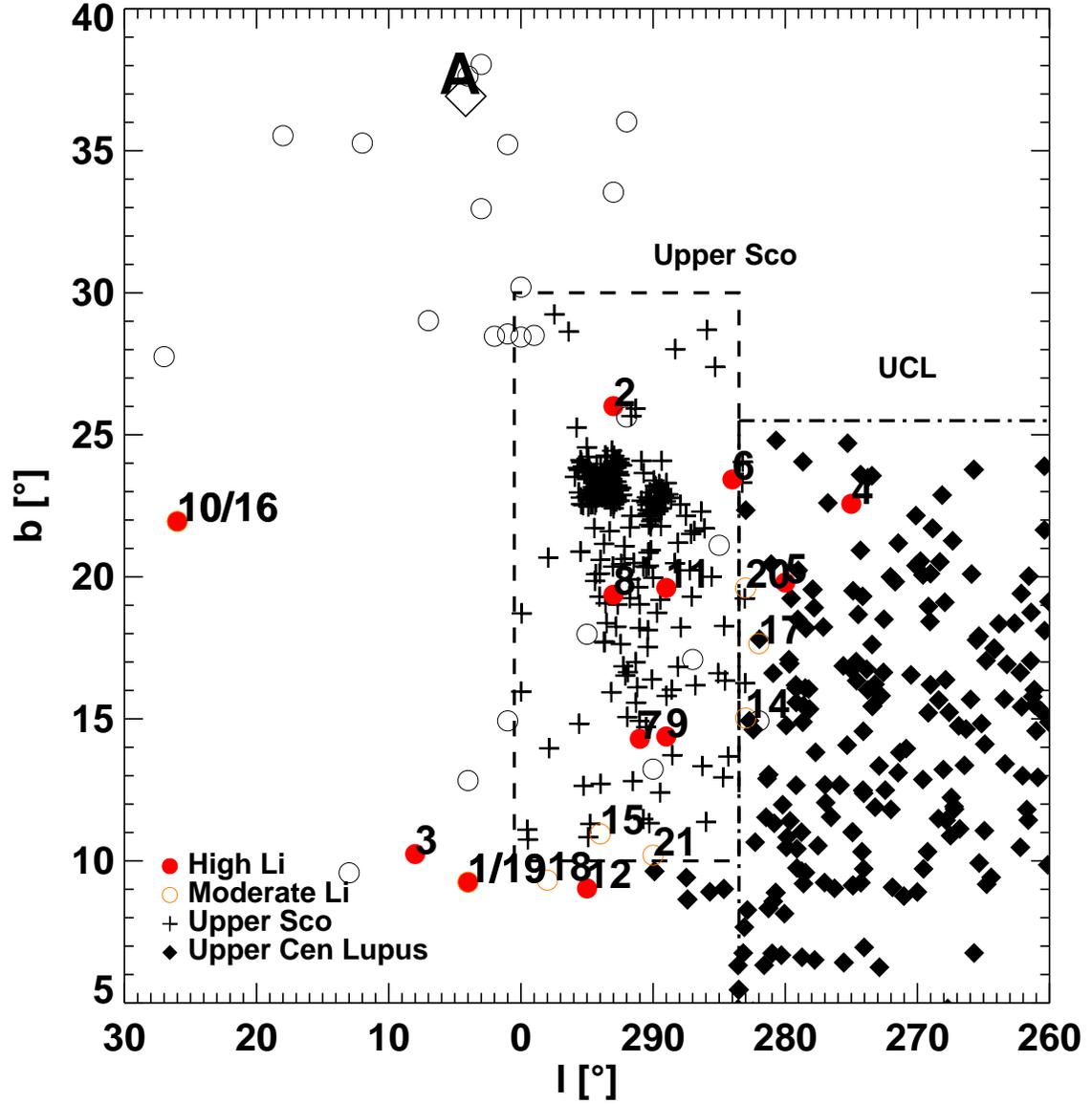}
\caption{\label{fig_zoomgc}
A cropped area about HD 141569 plotted with high and moderate lithium sample.  Symbols are defined 
as in Fig.\ \ref{fig_galcoords}.}
\end{figure}

Our procedure is rooted in the derivation of 
\citet[][see their $\S$2 and references therein]{deBruijne:1999}.  The process is executed 
assuming a velocity vector, and hence a convergent point, for the moving group to which we 
are testing membership.  In analyzing the spatial distribution of the 21 stars of our 
youthful sample, we note they are all in closer proximity to Upper Sco and UCL than HD 141569 (see   
Fig. \ref{fig_zoomgc}). 
The youthful sample is highly spatially separated from HD 141569, and these separations 
indicate two kinematic issues.  First, it is unlikely for objects with such large separations 
to be comoving.  Second, had these stars indeed formed together, large initial velocities 
($\sim$6 km s$^{-1}$, inconsistent with the observed 1-2 km s$^{-1}$ velocity dispersions of 
young associations) would be required to bring about the separations presently observed 
after $\sim$5 Myr of motion.  As it is unlikely these objects are associated with HD 141569, 
we require estimates of the mean velocity vectors for Upper Sco and UCL.  
The velocity vector for Upper Sco is adopted from \citet{Mamajek:2008}:  
$UVW$ = [$-$5.2,$-$16.6,$-$7.3] km s$^{-1}$. This vector incorporates a mean radial velocity 
for 120 Upper Sco members, an improvement over prior velocity vectors which solely rely upon 
proper motion and parallax information. For UCL, the velocity model used for comparison is 
derived from the median position, proper motion, and radial velocities of UCL members 
\citep{deZeeuw:1999}; $UVW$ = [$-$5.4,$-$19.7,$-$4.4] km s$^{-1}$.   
We calculate these $UVW$ vectors for Upper Sco and UCL to precisions of $\sim \pm$0.3 and 
$\sim \pm$0.4 km/s respectively, but note, there exist discrepancies between $UVW$ vectors 
derived by various authors.\footnote{For example, \citet{Madsen:2002} calculate for Upper Sco 
$UVW$ = [$-$0.9,$-$16.9,$-$5.3] km s$^{-1}$.  \citet{deBruijne:2001} cite 
$UVW$ = [4.1,$-$17.9,$-$3.7]  km s$^{-1}$ (with model-observation discrepancy parameter ``g'' 
set to equal nine) while \citet{deZeeuw:1999} report $UVW$ = [0.0,$-$16.1,$-$4.6] km s$^{-1}$.}  
The reason for the systematic differences between published 
convergent points and velocity vectors for the OB subgroups is not completely clear.
The leading candidates for these systematic differences are unaccounted-for expansion of 
the subgroups, and the probable presence of unresolved spatial and kinematic substructure 
within the subgroups.  For robustness, we include all available radial velocity measurements 
in derivation of UVW vectors.

For each of the 21 stars in our ``High Lithium'' and ``Moderate Lithium'' samples,
we derive moving cluster parallaxes using the HD 141569 and Upper Sco velocity
vectors (Table \ref{Astrometry}).  The formalism for this is:
\begin{equation}
\varpi = \frac{A \mu_{\upsilon} }{v \sin(\lambda)}
\end{equation}
where $\varpi$ is the parallax, A is 4.74 km yr s$^{-1}$ (the ratio of one AU in km to a Julian Year 
in s), $\mu_{\upsilon}$ is the parallel component of proper motion (proper motion in the direction 
of the convergent point), $v$ is the velocity of the group in km s$^{-1}$, and $\lambda$ is the 
angular separation between the star and the convergent point \citep[formula 1,][]{Mamajek:2005}.
An additionally useful parameter, the comovement probability, can also be calculated.  Comovement 
probability is defined as $1-P_\perp$, where $P_\perp$ is the likelihood that the star's proper motion 
is entirely perpendicular to the direction of the convergent point; the projection of proper motion 
in this direction is denoted $\mu_\tau$, and a $\mu_\tau$ close to 0 is indicative of comovement.

\section{RESULTS}\label{results}

\subsection{Distances, comovement probabilities, and membership}

The spatial proximity of our youthful sample stars to Upper Sco and UCL (Fig.\ \ref{fig_zoomgc}) 
suggests that these objects are not likely related kinematically to the farther away 
HD 141569 system.  It is important to stress at this juncture that derived comovement probabilities 
are not absolute probabilities per se; their derivation depends directly on the velocity model assumed 
{\em a priori}.  We do know with certainty that the stars in our sample are young (by virtue of their high 
Li abundances) and that they are moreover in projected proximity to other stars known to be young, nearby, 
and comoving (Fig.\ \ref{fig_galcoords}). Thus there is a strong ``prior'' favoring the velocity 
models that we have chosen to test. Still, the comovement probabilities reported in Table 
\ref{Astrometry} should be regarded as measures of {\it consistency} with the assumed velocity 
models, not proof of membership. We therefore adopt the very simple criterion of spatial 
proximity to a group in application of velocity modeling, and report the resulting distances 
and comovement probabilities for objects when tested against the velocity vectors of Upper Sco and UCL.
Parallax distances and comovement probabilities calculated as previously described (\S\ref{mcp}) 
are reported in Table \ref{Astrometry}.  

Based on two simple criteria, youth determined via measurement of the $\lambda$6707 line and spatial 
position, objects 2, 6, 7, 8, 9, 11, 13, 15, and 21 lie within the Upper Scorpius ``box'' as defined by 
\citet{deZeeuw:1999}.  Similarly, stars 4, 5, 14, 17, and 20 appear to be UCL members.  These ``spatial 
matches'' are summarized in column 3 of Table \ref{Membership}.  Due to the similarities of velocity vectors 
in the Ophiucus-Sco-Cen region, it is unsurprising that in some cases, objects we deem US or UCL members 
have higher comovement probabilities when tested against the velocity vector of the other group.  Factors 
which create blurring of kinematic boundaries include internal velocity dispersions inherent to a given 
moving group and observational uncertainties which then propagate into the convergent point solution.  
We thus take $(\ell,b)$ position as the strongest indicator of group membership and then examine comovement 
probabilities as a supplement.
Outside of the Upper Sco box, stars 3, 12, and 18 have high comovement probabilities with Upper Sco.  We 
would present 12 and 18 with some caution as Upper Sco members, as they are within a few degrees of the most 
extended, already known Upper Sco members.  

Object 3 is almost ten degrees away from the southernmost US stars and thus its association with Upper Sco 
is also dubious.  For these three objects, we tentatively suggest Upper Sco 
membership and denote their membership in Table \ref{Membership} as ``US?''  In two cases we note objects 
with low comovement probabilities with their spatially matched groups: objects 1 and 19 do not have velocities 
consistent with US and are spatially inconsistent with being UCL members.  Object 15, while spatially 
coincident with Upper Sco, has low enough comovement probability to be suspect.  Finally, stars 10 and 16 
have high comovement probabilities with Upper Sco, but appear to be too far away in $(\ell,b)$ space to be 
considered part of Upper Sco.  These remaining five objects we also classify as being of ``Indeterminate'' 
membership.  In summary, the total number of new Upper Sco members presented here is eight, and five new 
members of UCL are also identified.

\subsection{Space Motions}\label{sec_uvw}

To illustrate kinematic association in a familiar way, we could utilize the transformation 
matrices of \citet{Johnson:1987} to calculate $UVW$ space motions for the sample stars.  $UVW$ motions, 
however, depend on distance, a quantity we have obtained via assumption of comovement with a given 
$UVW$ vector.  The resulting $UVW$ plot is thus degenerate and does not provide additional criteria by 
which we can further examine association.

\begin{figure}[ht]
\plotone{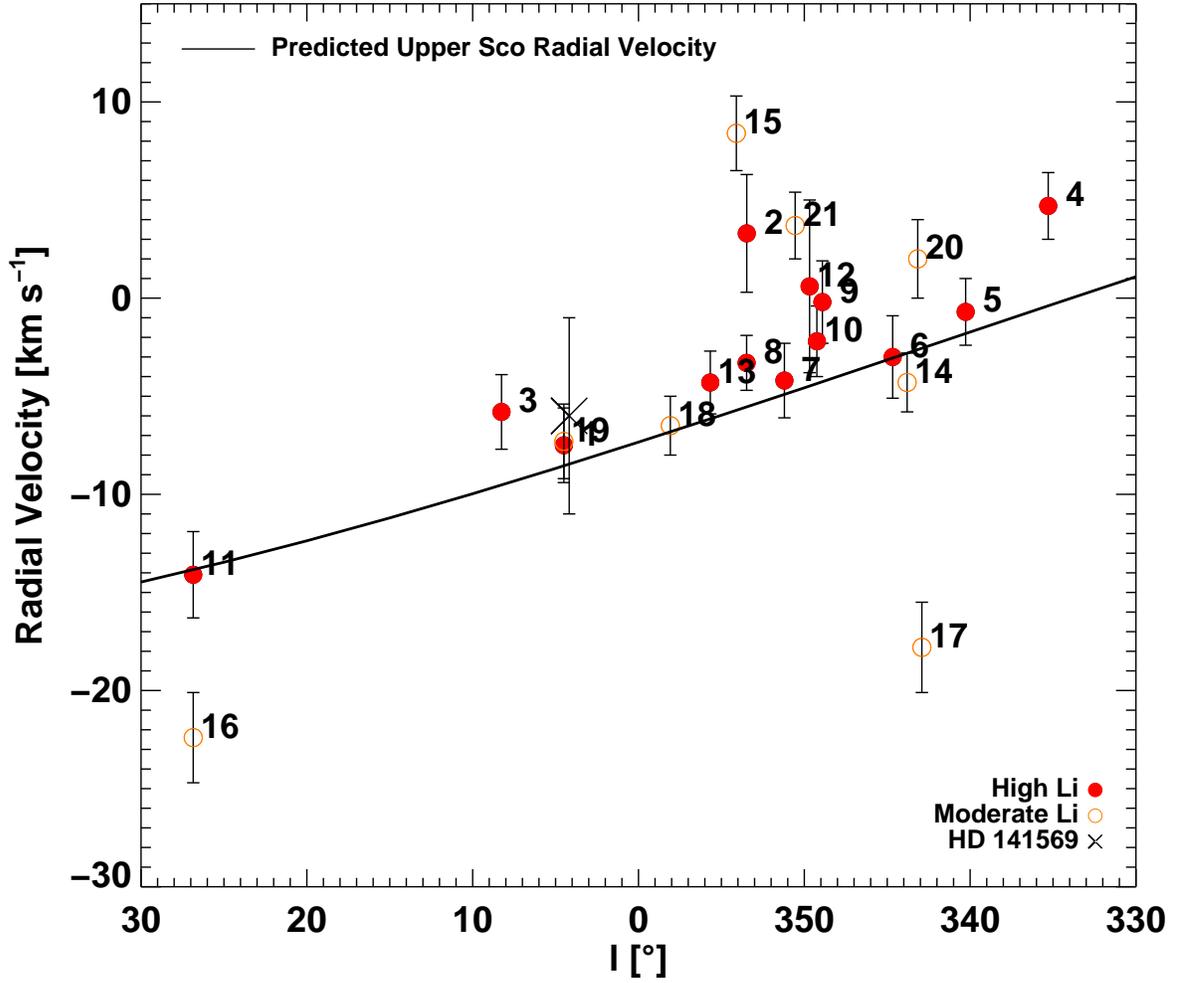}
\caption{\label{fig_radvelgal}
Radial velocities for the Lithium selected sample plotted as a function of galactic longitude.  
Also shown for comparison, the radial velocities predicted from the Upper Sco velocity vector 
at a galactic latitude of $+$20\degr are projected in galactic longitude.  Shown are 1$\sigma$ 
uncertainties; all high Lithium stars are consistent in radial velocity space with Upper Sco 
within 3$\sigma$.}
\end{figure}

As an additional check on the application of each velocity model, given an assumed velocity vector, 
radial velocities for each object can be predicted based on their proper motions and positions.  We 
find good consistency between the predicted radial velocities and those measured when comparing measured 
radial velocity to predicted radial velocity for whichever velocity model we would na\"{i}vely expect 
need be applied given simple spatial proximity to a given moving group.  Illustrating the radial velocity 
structure of the sample in context with nearby groups can be a measurement-based, assumption-free way of 
analyzing space motions.  In the selection criteria we constrained proper motions to agree with those of 
HD 141569 within a wide range of values which includes proper motions generally observed in Upper Sco.  
Measured radial velocities and projected radial velocities for the Upper Sco velocity vector are 
plotted as a function of galactic longitude in Fig. \ref{fig_radvelgal}.  Most notably, the entire 
``High Li'' sample agrees well with the predicted radial velocities of Upper Sco within $\sim$2-3$\sigma$.  
The farthest outlying points are from the ``Moderate Li'' sample.  

\subsection{H-R Diagram}\label{sec_hrd}

\begin{figure}[ht]
\plotone{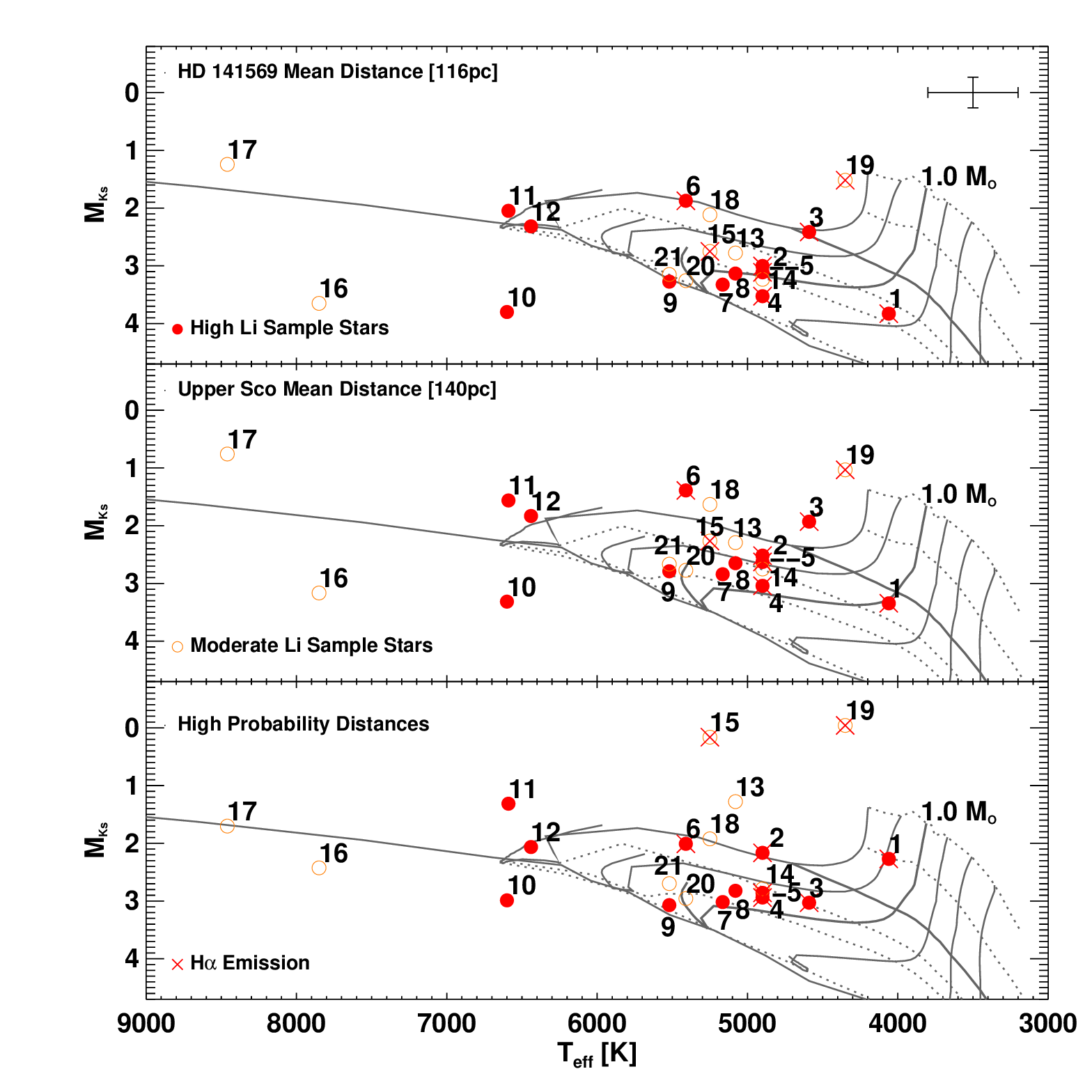}
\caption{\label{fig_HR}
HR Diagram for sample stars with the pre-main-sequence tracks of \citet{Baraffe:1998}
overplotted.  Isochrones shown are (from top to bottom) 1, 3, 10, 20, 30, and 100 Myr.
Also shown is the 100 Myr isochrone of \citet{Dantona:1997} extending to higher masses.
This isochrone and the 10 Myr isochrone are highlighted for reference.
The upper panel shows the stars using M$_{Ks}$ derived using the distance to 
HD 141569 from the second \textit{Hipparcos} data release, 116$\pm$8 pc \citep{vanLeeuwen:2007}.  
In the middle pane, we apply the mean distance to Upper Scorpius, 145$\pm$2 pc \citep{deZeeuw:1999}.  
The lower panel was generated using distances derived from the Upper Sco velocity model (see text). 
The sample's isochronal age appears consistent with the high lithium abundances which indicate ages 
$\lesssim$30 Myr (see Fig. \ref{fig_lithium}).}
\end{figure}

In Fig.\ \ref{fig_HR} we show the placement of the sample stars on three HR diagrams to illustrate 
shifts in M$_{Ks}$ magnitude due to changes in distance.  Absolute magnitudes were calculated 
from the observed 2MASS K$_S$ magnitudes (Table \ref{StellarParams}) and one of three distances.  
Uncertainties in M$_{Ks}$ are the propagated errors of the 2MASS photometry together with the formal errors 
in distance.  The uncertainty in $T_{\rm eff}$ is taken to be two spectral subtypes (see \S\ref{teff}).
To derive ages for our sample stars we also show the pre-main-sequence (PMS) evolutionary tracks of 
\citet{Baraffe:1998} and \citet{Dantona:1997}.

In the upper panel, we illustrate placement of the sample on the HRD when we apply 
the distance to the HD 141569 system to every individual object.  The isochronal ages inferred 
for most stars in our sample using the HD 141569 mean distance are in general older ($\sim 30$--100 Myr) 
than what would be expected for the stars based on their lithium abundances and comovement with the 
HD 141569 system (age $5\pm 3$ Myr). In concert with the lack of spatial proximity, we further rule out 
the potential for a coeval, coherent moving group near HD 141569.

In contrast, the inferred ages using the Upper Sco mean distance ($\sim$145pc, effectively 
equivalent that of UCL, $\sim$142pc) are entirely consistent with the expectation of 
$\lesssim$30 Myr as imposed by the Li EW measurements.  All stars appear on or above the 30 Myr
isochrone, save objects \#10 and \#16, which, despite their high comovement probabilities with 
Upper Sco, do not appear to be in close enough spatial proximity to be members of that moving group.
In the third HR diagram, applied distances are determined by which velocity model, US or UCL, 
provides the highest comovement probability with a given object.  This particular representation is 
not only mostly consistent with the age range expected from Li {\sc I} presence, it also 
``correctly'' places the higher mass objects closer to the ZAMS; particularly, objects 10, 16, and 17 
have derived distances which make them appear to be ZAMS stars rather than anomalous objects far above 
or below the theoretical isochrones.

Scatter in an HRD generally can be attributed to many factors; the radial extent of a young association 
\citep[][e.g., TW Hydra $\sim$55pc]{Mamajek:2005}, observational errors, or even the choice of 
evolutionary tracks can generate shifts and enhance spread in isochronal age of a sample expected to be 
coeval.  Using the \citet{Dantona:1997} evolutionary models, our sample stars all appear to have 
isochronal ages $\lesssim$10 Myr.  In general, these tracks appear shifted by $\sim$400K to higher 
temperatures with respect to the \citet{Baraffe:1998} tracks.  For an illustration of these track-based 
discrepancies, see \citet{Simon:2000}.  In spite of obstacles posed by apparently discrepant HR diagrams, 
we can say with confidence based on Li {\sc I} presence that these objects are indeed young, $\lesssim$30 Myr.
The HR diagram, when applying high comovement probability derived distances, provides a higher degree 
of confidence in adopting these distances as the ages are indeed as expected from Li measurements.

\subsection{Is HD 141569 Related to Upper Sco?}\label{sec_hdusdisc}

HD 141569 is an apparently isolated system located within tens of degrees (and parsecs) of known sites 
of recent ($<$5 Myr) and ongoing star formation, all apparently associated with the Sco-Cen star forming 
complex ($d$ = 100-200 pc; Preibisch \& Mamajek, 2008, in press), which appears contiguous with the 
Aquila Rift regions \citep[][$\ell$ $\simeq$ 30$\degr$;]{Dame:1987}.  HD 141569 appears to agree with 
the projected velocity model of Upper Sco (see Fig. \ref{fig_radvelgal}, but we can rule out the possibility 
of HD 141569 originating or being kinematically associated with Upper Sco.  The hypothesis that HD 141569 
could have been ejected at high velocity from a known high density stellar nursery can be strongly 
discounted on two grounds.  First, HD 141569 has two low-mass companions at wide separation 
($\sim$10$^3$ AU, with likely orbital motion of $\sim$1 km s$^{-1}$). A velocity kick of $>$2-3 km s$^{-1}$ 
to either A or B+C would have likely disintegrated the system. A low velocity ejection ($<$2-3 km s$^{-1}$) 
would have placed the birth site within $<$10-15 pc ($<$5$^{\circ}$-7$^{\circ}$), but no such known young 
clusters or molecular clouds appear there.  


Position and velocity information for these nearby groups was entered into an orbit code which employs 
the epicyclic approximation. The separations between these groups and HD 141569 were evaluated during 
the past 10 Myr; combined distance and velocity vector uncertainties result in $<$15 pc uncertainties 
over this time frame.  Presently, HD 141569 is $\sim$55 pc away from the center of Upper Sco, and was 
only slightly closer at its minimum separation of $\sim$53 pc ($\sim$2.7 Myr ago).  In $UVW$, the only 
substantial difference is in the $W$ component of velocity: while Upper Sco has negligible vertical 
motion with respect to the Local Standard of Rest \citep{Mamajek:2008}, HD 141569 is moving northward 
out of the disk at $\sim$5 km s$^{-1}$.  This anomalous $W$ component of motion is discrepant with any 
known molecular cloud or star forming region near HD 141569.  Further, given the kinematic data and the 
isochronal age of HD 141569 ($\sim$4-5 Myr), it appears that HD 141569 {\it could not have formed from 
any of the known sites of recent star-formation in its vicinity}. The list of excluded birth-sites includes 
Upper Sco, UCL, Lower Centaurus Crux (LCC), and the Ophiuchus, Corona Australis, and Lupus clouds.

Combining the kinematic, position, and age data, we conclude that the HD 141569 triple system likely 
formed in isolation or with a small entourage of companions in a cloud. It appears to have formed $\sim$25 pc 
closer to the Galactic plane than its present position, and its anomalously large W velocity component has carried 
it to its modern high latitude position ($b$ $\simeq$ +37$\degr$, $\sim$90 pc above the Galactic plane).

\section{Summary and Conclusions}\label{summconc}

We have identified a group of 21 PMS stars within 30\degr\ of HD 141569 on the basis of strong 
Li {\sc I} absorption and, in 9 of those 21 cases, H$\alpha$ in emission. These stars were selected 
through a joint catalog search for X-ray sources with spatial and proper-motion characteristics 
similar to those of HD 141569, a B9.5Ve star at 116 pc that harbors a circumstellar disk and for which 
two low-mass companions had previously been identified \citep{Weinberger:2000}.  For these 21 stars, we 
have applied a moving cluster parallax technique to proper-motion data from the literature.  

Table \ref{Membership} outlines our final membership assessments: we present eight potential new members 
of Upper Sco and five new potential members of UCL.  These stars possess lithium presence consistent 
with youth and futhermore appear youthful on the HR diagram.  Primarily we utilize spatial position as the 
principal criterion for determining membership and supplement that with comovement probabilties from the 
moving cluster parallax derivation.  Additionally, we examine the motion of the HD 141569 system away from 
the galactic midplane over its lifetime.  The system, surprisingly, appears to have formed in isolation, 
well outside of presently known star forming regions and molecular clouds.  

\acknowledgments
The authors would like to acknowledge the support of a NASA Origins 
of Solar Systems grant to A.~J.~W., the NSF Research Experience for 
Undergraduates program, and the NASA Astrobiology Institute.  This 
work was also made possible by an NSF Career award to K.~G.~S. (AST-0349075).  
E.~E.~M. acknowledges support through a Clay Postdoctoral Fellowship from 
Smithsonian Astrophysical Observatory.  The authors also note that the Vizier 
search engine \citep{Ochsenbein:2000} was invaluable in this project.  We 
thank Drs. Suzan Edwards, Eric Jensen, and Inseok Song as well as an 
anonymous referee for assistance and input during the analysis and 
revision processes.

\clearpage

\begin{landscape}
\begin{deluxetable}{lcccccc}
\tabletypesize{\scriptsize}
\tablecaption{Observing Log\label{obslog}}
\tablewidth{0pt}
\tablehead{\colhead{Object Name} & \colhead{Right Ascension} & \colhead{Declination} & \colhead{Observation Time}   & \colhead{Integration}  & \colhead{S/N\tablenotemark{(1)}} & \colhead{Comment(s)} \\
           \colhead{}            & \colhead{[J2000]}         & \colhead{[J2000]}     & \colhead{[UT]}               & \colhead{Time [s]}     & \colhead{}                       & \colhead{}}
\startdata
\multicolumn{1}{c}{2001-06-18 : UCO / Lick}\\ \hline
Alpha Boo               &	14:15:39.67	&	+19:10:56.7     &       04:17:13.0      &	    1	&	200	&	K2III $v_{r}$ Standard		\\
HD 137396               &	15:26:05.91	&	-11:41:55.7  	&	04:28:35.0      &	  720	&	127	&	\nodata				\\
RHS 48   	        &	15:23:46.0	&	-00:44:25    	&	04:52:38.0      &	 1500	&	158	&	\nodata				\\
HD 138969               &	15:35:47.41	&	-12:51:32.9  	&	05:25:23.0      &	  720	&	105	&	\nodata				\\
HD 140574               &	15:44:26.30	&	-03:50:18.5  	&	05:45:22.0      &	  720	&	132	&	\nodata				\\
HD 141612               &	15:50:11.35	&	-06:32:14.6  	&	06:05:08.0      &	 1500	&	178	&	\nodata				\\
HD 141693               &	15:50:11.86	&	+05:57:17.2  	&	06:37:21.0      &	  360	&	105	&	\nodata				\\
HD 142987 	        &	15:58:20.558	&	-18:37:25.09 	&	06:50:42.0      &	 1800	&	75	&	\nodata				\\
TYC 6191-0552           &	15:58:47.724	&	-17:57:59.79 	&	07:30:24.0      &	 2400	&	31	&	\nodata				\\
HD 143332               &	16:00:02.65	&	-07:07:21.8  	&	08:17:18.0      &	  600	&	126	&	\nodata				\\
HIP 79354               &	16:11:44.57	&	+00:47:38.6  	&	08:34:18.0      &	 1000	&	97	&	\nodata				\\
HD 145551               &	16:11:56.93	&	-10:37:20.3  	&	08:58:01.0      &	  600	&	98	&	\nodata				\\
BD -06 4414 	        &	16:22:17.866	&	-06:23:03.93 	&	09:15:34.0      &	  600	&	72	&	\nodata				\\
HD 143810               &	16:02:47.47	&	-10:37:56.7  	&	09:32:25.0      &	  600	&	72	&	\nodata				\\
HD 144726               &	16:07:44.91	&	-11:55:13.1  	&	09:49:16.0      &	  600	&	145	&	\nodata				\\
HD 145169               &	16:09:50.80	&	-11:07:56.5  	&	10:06:16.0      &	  450	&	95	&	\nodata				\\
HD 177178               &       19:03:32.25     &       +01:49:07.6     &	10:20:40.0      &	  300	&	175	&	A4V $v_{r}$ Standard		\\
HD 187691               &	19:51:01.64	&	+10:24:56.6  	&	10:33:26.0      & 	  240	&	184	&	F8V $v_{r}$ Standard		\\
16 Cyg/HD186427         &       19:40:32.06     &       +50:31:03.1     &	10:45:50.0      &	  240	&	197	&	G2V $v_{r}$ Standard		\\
HD 184467               &       19:31:07.97     &       +58:35:09.6     &	10:59:00.0      &	  300	&	139	&	K2V $v_{r}$ Standard		\\ \hline
\multicolumn{1}{c}{2002-04-17 : LCO}\\ \hline
HD 80170                &       09:16:57.08     &       -39:24:00.5     &	01:07:53.4      &	  30	&	107	&	K5III $v_{r}$ Standard		\\
HD 102870               &       11:50:41.72     &       +01:45:53.0     &	01:16:31.1      &	  20	&	131	&	F9V $v_{r}$ Standard		\\
TYC 6141-0525-1         &	14:06:06.475	&	-18:10:37.41 	&	03:17:19.9      &	1080	&	81	&	\nodata				\\
TYC 0909-0125-1         &	14:19:33.692	&	+08:38:09.59 	&	03:43:18.4      &	1080	&	79	&	\nodata				\\
TYC 7312-0236-1         &	15:12:44.473	&	-31:16:48.10 	&	00:00:00.0      &	1500	&	89	&	\nodata				\\
TYC 7327-0689-1         &	15:37:51.344	&	-30:45:16.15 	&	04:37:09.7      &	1500	&	77	&	\nodata				\\
TYC 5003-0138-1         &	15:26:52.727	&	-00:53:11.74 	&	05:09:45.3      &	1500	&	112	&	\nodata				\\
TYC 0937-0754-1         &	15:38:00.94	&	+11:34:57.30 	&	05:41:42.1      &	1500	&	75	&	\nodata				\\
TYC 6781-0415-1         &	15:41:31.212	&	-25:20:36.39 	&	06:16:54.2      &	 900	&	86	&	\nodata				\\
TYC 6790-1227-1         &	15:48:02.922	&	-29:08:37.02 	&	06:40:12.4      &	1500	&	101	&	\nodata				\\
HD 142016               &	15:53:12.969	&	-30:46:44.17 	&	07:26:45.0      &	 255	&	58	&	\nodata				\\
TYC 0376-0769-1         &	16:19:59.977	&	+04:36:46.33 	&	07:42:44.8      &	1200	&	85	&	\nodata				\\
TYC 7346-1182-1         &	16:42:07.711	&	-30:38:37.87 	&	08:12:47.3      &	1200	&	117	&	\nodata				\\
HD 153439               &	17:00:42.963	&	-27:25:17.55 	&	08:37:23.9      &	 750	&	136	&	\nodata				\\
CD -25 11942             &	17:06:01.190	&	-25:20:30.40 	&	08:57:16.9      &	 900	&	150	&	\nodata				\\
TYC 6817-1757-1         &	16:49:35.996	&	-27:28:07.78 	&	09:18:32.5      &	1350	&	80	&	\nodata				\\
HD 188376               &	19:55:50.36	&	-26:17:58.2  	&	09:47:46.4      &	  65	&	109	&	G5V $v_{r}$ Standard		\\
HD 165341               &	18:05:27.29	&	+02:30:00.4  	&	09:58:15.0      &	  45	&	206	&	K0V Spectral Type Standard	\\
HD 209290	        &	22:02:10.27	&	01:24:00.8   	&	10:06:48.1	&	 750	&	52	&	M0.5V Spectral Type Standard	\\ \hline
\multicolumn{1}{c}{2002-04-18 : LCO}\\ \hline
TYC 5022-0263-1         &	15:48:40.942	&	-03:10:44.22 	&	06:22:52.4      &	1500	&	116	&	\nodata				\\
HD 143358	        &	16:01:07.927	&	-32:54:52.53 	&	06:54:12.0	&	 900	&	117	&	\nodata				\\
HD 144713               &	16:08:05.223	&	-24:55:33.2  	&	07:14:23.6      &	 900	&	153	&	\nodata				\\
HD 144732               &	16:08:26.317	&	-28:25:49.3  	& 	07:34:59.7      &	1200	&	105	&	\nodata				\\
TYC 6806-0888-1         &	16:23:46.978	&	-28:50:02.43 	&	07:59:41.7      &	1800	&	100	&	\nodata				\\
TYC 6803-0897-1         &	16:29:49.911	&	-27:28:49.96 	&	08:35:31.7      &	1800	&	 99	&	\nodata				\\
HD 148982               &	16:32:50.876	&	-28:20:39.89 	&	09:11:17.0      &	1200	&	115	&	\nodata				\\
HD 157310               &	17:22:23.477	&	+04:45:24.83 	&	09:38:10.9      &	1200	&	 99	&	Double star - brighter component\\
BD $+$04 3405B          &	17:22:23.5	&	+04:45:13.3  	&	09:59:37.2      &	1200	&	 60	&	Double star - fainter component \\ \hline
\multicolumn{1}{c}{2002-04-19 : LCO}\\ \hline
HD 141813        	&	15:51:54.393	&	-26:22:05.52 	&	05:41:33.5      &	 900	&	120	&	\nodata				\\
HD 148396        	&       16:28:51.327    &       -30:16:55.26    &	06:06:32.7      &	1500	&	119	&	Catalogued Double star		\\
TYC 0976-1617-1  	&	17:01:57.543	&	+07:33:32.45 	&	06:40:26.6      &	2400	&	100	&	SB $^{(2)}$			\\
HD 154922        	&	17:09:07.521	&	-18:15:22.78 	&	07:27:02.2      &	 450	&	107	&	\nodata				\\
2MASS J17215666-2010498	&	17:21:56.67	&	-20:10:49.8     &	08:03:30.1      &	3600	&	52	&	Double star - brighter component\\
TYC 6242-0104-1         &	17:21:56.048	&	-20:10:51.97 	&	09:09:59.6      &	3600	&	71	&	Double star - fainter component \\ \hline
\multicolumn{1}{c}{2002-04-20 : LCO}\\ \hline
HD 109524	        &	12:35:33.55	&	-34:52:54.9  	&	04:11:22.8	&	 750	&	238	&	K2V $v_{r}$ Standard	        \\
HIP 75685               &	15:27:42.638	&	-02:45:18.56 	&	05:43:37.4      &	 900	&	122	&	\nodata			        \\
TYC 6234-1287-1         &	17:26:56.545	&	-16:31:34.85 	&	06:06:12.0      &	1800	&	118	&	\nodata			        \\
TYC 7334-0429-1         &	16:04:30.557	&	-32:07:28.7  	& 	06:43:18.0      &	1500	&	110	&	\nodata			        \\
TYC 5668-0365-1         &	17:40:55.47	&	-12:16:27.75 	&	07:14:42.1      &	2400	&	134	&	\nodata			        \\
HD 144393               &	16:06:06.470	&	-12:18:15.22 	&	08:01:37.6      &	 360	&	109	&	\nodata			        \\
TYC 6214-2384-1         &	16:19:33.958	&	-22:28:29.41 	&	08:59:30.1      &	1800	&	99	&	\nodata			        \\
TYC 6215-0184-1         &	16:29:48.700	&	-21:52:11.89 	&	09:35:25.5      &	2400	&	129	&	SB $^{(2)}$		        \\
HD 177178     	        &	19:03:32.25	&	+01:49:07.6  	&	10:24:04.7      &	 150	&	109	&	A4V $v_{r}$ Standard	        \\
\enddata
\tablecomments{}
\tablenotetext{1.}{Approximated using \textit{Splot} at $\sim$6500\AA and $\sim$6700\AA.}
\tablenotetext{2.}{Suspected spectroscopic binary based upon broadened troughs of spectral features.}
\end{deluxetable}
\end{landscape}

\begin{deluxetable}{llcccccc}
\tabletypesize{\scriptsize}
\tablecaption{Stellar Parameters\label{StellarParams}}
\tablewidth{0pt}
\tablehead{\colhead{Plot} & \colhead{Object Name} & \colhead{$J$\tablenotemark{(1)}}   & \colhead{$H$\tablenotemark{(1)}} & \colhead{$K_{s}$\tablenotemark{(1)}} & \colhead{$\mu_{\alpha}$\tablenotemark{(2)}} & \colhead{$\mu_{\delta}$\tablenotemark{(2)}}   & \colhead{Parallax\tablenotemark{(3)}} \\
           \colhead{ID}   & \colhead{}            & \colhead{}                         & \colhead{}                       & \colhead{}                           & \colhead{[mas yr$^{-1}$]}                  & \colhead{[mas yr$^{-1}$]}                    & \colhead{[mas]}    }
\startdata
 A  & HD 141569        &        6.872$\pm$0.027        &       6.861$\pm$0.040 &       6.281$\pm$0.026 &         -18.3$\pm$1.1 &       -20.5$\pm$1.1           &       8.63$\pm$0.59   \\ \hline
 1  & TYC 6242-0104-1  &        9.963$\pm$0.027        &       9.305$\pm$0.026 &       9.151$\pm$0.024 &         -11.7$\pm$3.6 &       -13.7$\pm$4.0           &       \nodata         \\
 2  & TYC 6191-0552    &        9.261$\pm$0.022        &       8.535$\pm$0.042 &       8.325$\pm$0.024 &         -15.0$\pm$3.3 &       -20.2$\pm$3.7           &       \nodata         \\
 3  & TYC 6234-1287-1  &        8.659$\pm$0.025        &       8.012$\pm$0.024 &       7.829$\pm$0.020 &         -10.1$\pm$2.6 &       -39.0$\pm$2.7           &       \nodata         \\
 4  & TYC 7312-0236-1  &        9.601$\pm$0.024        &       9.079$\pm$0.024 &       8.919$\pm$0.019 &         -20.5$\pm$3.3 &       -20.5$\pm$3.2           &       \nodata         \\
 5  & TYC 7327-0689-1  &        9.300$\pm$0.024        &       8.755$\pm$0.036 &       8.563$\pm$0.019 &         -19.7$\pm$3.2 &       -27.5$\pm$3.1           &       \nodata         \\
 6  & TYC 6781-0415-1  &        7.974$\pm$0.030        &       7.367$\pm$0.033 &       7.241$\pm$0.024 &         -19.5$\pm$2.7 &       -30.1$\pm$2.4           &       \nodata         \\
 7  & TYC 6803-0897-1  &        9.275$\pm$0.024        &       8.743$\pm$0.049 &       8.648$\pm$0.025 &         -15.6$\pm$2.5 &       -28.6$\pm$2.5           &       \nodata         \\
 8  & TYC 6214-2384-1  &        9.230$\pm$0.019        &       8.659$\pm$0.036 &       8.509$\pm$0.019 &         -18.7$\pm$3.5 &       -26.2$\pm$3.8           &       \nodata         \\
 9  & TYC 6806-0888-1  &        9.216$\pm$0.025        &       8.785$\pm$0.027 &       8.659$\pm$0.026 &         -13.4$\pm$3.0 &       -27.5$\pm$2.7           &       \nodata         \\
10  & BD $+$04 3405B   &        9.759$\pm$0.022        &       9.413$\pm$0.031 &       9.262$\pm$0.019 &          -5.3$\pm$1.5*&       -14.9$\pm$1.8*          &       \nodata         \\
11  & HD 144713        &        7.847$\pm$0.021        &       7.538$\pm$0.034 &       7.431$\pm$0.020 &         -11.7$\pm$1.7 &       -20.7$\pm$1.6           &       \nodata         \\
12  & HD 153439        &        8.073$\pm$0.020        &       7.852$\pm$0.049 &       7.729$\pm$0.047 &          -6.7$\pm$1.6 &       -28.6$\pm$1.6           &       \nodata         \\ \hline 

13  & HD 148396        &        8.420$\pm$0.023        &       8.095$\pm$0.019 &       8.100$\pm$0.020 &          -6.9$\pm$2.4 &       -15.4$\pm$2.4           &       \nodata         \\
14  & TYC 7334-0429-1  &        9.168$\pm$0.018        &       8.690$\pm$0.049 &       8.565$\pm$0.021 &         -17.5$\pm$2.2 &       -25.5$\pm$2.2           &       \nodata         \\
15  & TYC 6817-1757-1  &        8.815$\pm$0.021        &       8.350$\pm$0.042 &       8.179$\pm$0.031 &         -10.1$\pm$2.8 &       -7.4$\pm$2.5            &       \nodata         \\
16  & HD 157310        &        9.160$\pm$0.022        &       9.088$\pm$0.047 &       9.006$\pm$0.021 &          -5.4$\pm$1.5 &       -12.0$\pm$1.5           &       \nodata         \\
17  & HD 142016        &        6.785$\pm$0.020        &       6.707$\pm$0.034 &       6.622$\pm$0.018 &         -26.4$\pm$1.2 &       -38.8$\pm$1.3           &       \nodata         \\
18  & CD -25 11942     &        8.099$\pm$0.020        &       7.661$\pm$0.029 &       7.525$\pm$0.038 &          -9.8$\pm$2.0 &       -28.2$\pm$1.8           &       \nodata         \\
19  & 2MASS J17215666-2010498 &        8.150$\pm$0.023        &       7.187$\pm$0.047 &       6.840$\pm$0.023 &            \nodata    &           \nodata             &       \nodata         \\
20  & TYC 6790-1227-1  &        9.212$\pm$0.023        &       8.719$\pm$0.026 &       8.624$\pm$0.023 &         -20.4$\pm$2.8 &       -26.0$\pm$2.3           &       \nodata         \\
21  & TYC 7346-1182-1  &        9.018$\pm$0.027        &       8.663$\pm$0.053 &       8.530$\pm$0.019 &         -14.3$\pm$2.3 &       -27.0$\pm$2.2           &       \nodata         \\ \hline

  & HD 142987       &        8.279$\pm$0.035        &       7.774$\pm$0.063 &       7.614$\pm$0.021 &         -15.4$\pm$2.1 &       -22.3$\pm$2.2           &       \nodata         \\
  & HD 143358       &        8.470$\pm$0.023        &       8.164$\pm$0.036 &       8.074$\pm$0.020 &        -18.3$\pm$1.4  &       -29.6$\pm$1.5           &       \nodata         \\
  & BD -06 4414     &        8.359$\pm$0.027        &       8.043$\pm$0.038 &       7.936$\pm$0.033 &         -27.0$\pm$2.1 &       -33.7$\pm$2.2           &       \nodata         \\
  & HD 137396       &        7.632$\pm$0.023        &       7.493$\pm$0.033 &       7.419$\pm$0.027 &       -14.9$\pm$1.4   &       -12.0$\pm$1.0           &       \nodata         \\
  & HD 138969       &        7.946$\pm$0.027        &       7.693$\pm$0.040 &       7.666$\pm$0.017 &       -11.3$\pm$1.4   &       -8.5$\pm$1.0            &       10.69$\pm$1.4   \\
  & HD 140574       &        7.623$\pm$0.018        &       7.495$\pm$0.036 &       7.459$\pm$0.029 &       -27.2$\pm$1.3   &       -30.2$\pm$0.9           &       10.8$\pm$1.15   \\
  & HD 141612       &        8.808$\pm$0.024        &       8.497$\pm$0.057 &       8.398$\pm$0.031 &       -17.7$\pm$2.4   &       -10.7$\pm$1.8           &       9.62$\pm$1.74   \\
  & HD 141693       &        6.886$\pm$0.020        &       6.888$\pm$0.034 &       6.828$\pm$0.023 &       -29.7$\pm$1.2   &       -26.5$\pm$0.7           &       8.96$\pm$0.96   \\
  & HD 141813       &        8.232$\pm$0.023        &       7.963$\pm$0.036 &       7.862$\pm$0.020 &         -22.7$\pm$1.7 &       -38.1$\pm$1.9           &       \nodata         \\
  & HD 143332       &        6.987$\pm$0.024        &       6.783$\pm$0.044 &       6.680$\pm$0.023 &         -10.6$\pm$1.3 &       -24.9$\pm$1.3           &       9.96$\pm$1.36   \\
  & HD 143810       &        8.953$\pm$0.030        &       8.766$\pm$0.061 &       8.643$\pm$0.019 &         -24.8$\pm$1.5 &       -13.7$\pm$1.5           &       9.26$\pm$1.51   \\
  & HD 144393       &        7.323$\pm$0.027        &       7.099$\pm$0.046 &       6.980$\pm$0.021 &          -6.3$\pm$1.2 &       -23.9$\pm$1.1           &       10.97$\pm$1.14  \\
  & HD 144726       &        7.527$\pm$0.027        &       7.322$\pm$0.036 &       7.244$\pm$0.026 &         -16.7$\pm$1.4 &       -19.4$\pm$1.3           &       8.49$\pm$1.20   \\
  & HD 144732       &        8.531$\pm$0.023        &       8.231$\pm$0.051 &       8.147$\pm$0.026 &         -15.2$\pm$2.1 &       -33.2$\pm$1.9           &       \nodata         \\
  & HD 145169       &        7.159$\pm$0.039        &       6.912$\pm$0.040 &       6.876$\pm$0.023 &         -20.2$\pm$1.1 &       -30.5$\pm$1.1           &       12.75$\pm$1.78  \\
  & HD 145551       &        7.906$\pm$0.029        &       7.677$\pm$0.042 &       7.600$\pm$0.018 &         -20.3$\pm$1.3 &       -19.4$\pm$1.3           &       9.48$\pm$1.51   \\
  & HD 148982       &        8.669$\pm$0.026        &       8.381$\pm$0.036 &       8.308$\pm$0.023 &         -15.8$\pm$2.0 &       -25.4$\pm$1.7           &       \nodata         \\
  & HD 154922       &        7.742$\pm$0.023        &       7.540$\pm$0.042 &       7.456$\pm$0.021 &          -7.4$\pm$1.3 &       -13.1$\pm$1.2           &       \nodata         \\
  & HIP 75685       &        9.186$\pm$0.024        &       8.870$\pm$0.042 &       8.810$\pm$0.024 &         -28.8$\pm$1.5 &       -19.4$\pm$1.4           &       8.92$\pm$1.72   \\
  & HIP 79354       &        8.090$\pm$0.023        &       7.752$\pm$0.031 &       7.649$\pm$0.021 &         -12.1$\pm$1.0 &       -29.8$\pm$1.0           &       9.18$\pm$1.99   \\
  & RHS 48          &        8.465$\pm$0.029        &       7.830$\pm$0.053 &       7.624$\pm$0.024 &         -15.1$\pm$1.6 &       -10.6$\pm$1.7           &       \nodata         \\
  & TYC 0376-0769-1 &        8.196$\pm$0.024        &       7.649$\pm$0.031 &       7.496$\pm$0.036 &         -18.8$\pm$1.8 &       -4.4$\pm$1.7            &       \nodata         \\
  & TYC 0909-0125-1 &        9.433$\pm$0.026        &       8.987$\pm$0.022 &       8.896$\pm$0.023 &         -16.2$\pm$2.7 &       -17.6$\pm$2.6           &       \nodata         \\
  & TYC 0937-0754-1 &        9.016$\pm$0.029        &       8.525$\pm$0.040 &       8.404$\pm$0.023 &         -13.1$\pm$2.0 &       -1.4$\pm$2.0            &       \nodata         \\
  & TYC 0976-1617-1 &       10.158$\pm$0.026        &       9.798$\pm$0.026 &       9.664$\pm$0.021 &         -11.0$\pm$3.0 &       -15.1$\pm$3.1           &       \nodata         \\
  & TYC 5003-0138-1 &        9.084$\pm$0.023        &       8.492$\pm$0.027 &       8.317$\pm$0.026 &         -27.3$\pm$1.6 &       -13.1$\pm$1.5           &       \nodata         \\
  & TYC 5022-0263-1 &        8.855$\pm$0.029        &       8.401$\pm$0.047 &       8.244$\pm$0.031 &          -6.7$\pm$3.2 &       -24.5$\pm$3.4           &       \nodata         \\
  & TYC 5668-0365-1 &        9.564$\pm$0.024        &       9.033$\pm$0.026 &       8.840$\pm$0.025 &          -2.8$\pm$2.2 &       -18.7$\pm$2.3           &       \nodata         \\
  & TYC 6141-0525-1 &        9.335$\pm$0.027        &       8.957$\pm$0.024 &       8.872$\pm$0.024 &         -15.2$\pm$2.3 &       -35.9$\pm$2.4           &       \nodata         \\
  & TYC 6215-0184-1 &        8.677$\pm$0.026        &       8.003$\pm$0.036 &       7.756$\pm$0.024 &          -3.6$\pm$2.9 &       -21.8$\pm$3.1           &       \nodata         \\
\enddata
\tablecomments{For star \#19, we adopt proper motions of its companion, \#1.}
\tablenotetext{1.}{From 2MASS Catalog.}
\tablenotetext{2.}{Tycho-2 proper motions.}
\tablenotetext{3.}{Hipparcos parallaxes.}
\tablenotetext{*}{Proper motions from UCAC2.}
\end{deluxetable}

\begin{landscape}
\begin{deluxetable}{llcccc|ccc|ccc|c}
\tabletypesize{\scriptsize}
\tablecaption{Effective Temperatures and Lithium Equivalent Widths\label{Spectroscopy}}
\tablewidth{0pt}
\tablehead{\colhead{Plot} & \colhead{Object Name} & \colhead{Li {\sc I} EW}  & \colhead{Li {\sc I} EW}  & \colhead{Ctmn.}   & \colhead{H$\alpha$   }            & \colhead{Spectral}  & \colhead{Type }                      & \colhead{T$_{\rm eff}$} & \colhead{$\lambda$6200/$\lambda$6210} & \colhead{Spectral} & \colhead{T$_{\rm eff}$} & \colhead{Adopted T$_{\rm eff}$}    \\
           \colhead{ID}   & \colhead{}            & \colhead{[m\AA] Integ.}   & \colhead{[m\AA] GFit}     & \colhead{[m\AA]}  & \colhead{Flag\tablenotemark{(1)}} & \colhead{Type}      & \colhead{Source\tablenotemark{(2)}}  & \colhead{[K]}          & \colhead{Line Ratio}                  & \colhead{Type}     & \colhead{[K]}          & \colhead{[K]}}
\startdata
 1 &  TYC 6242-0104-1  &  491     &       487     & 25      & e*   &       K5 Ve     &       2       &       4350    &       1.07    &       K5      &       4350     &  4350  \\
 2 &  TYC 6191-0552    &  481     &       492     & 15      & e*   &       K2        &       1       &       4900    &       1.91    &       K2      &       4900     &  4900  \\
 3 &  TYC 6234-1287-1  &  464     &       452     & 18      & e*   &       K4 Ve     &       2       &       4590    &       2.65    &       K1.5    &       4990     &  4590  \\
 4 &  TYC 7312-0236-1  &  434     &       433     & 15      & e*   &       K2 Ve     &       2       &       4900    &       2.30    &       K2      &       4900     &  4900  \\
 5 &  TYC 7327-0689-1  &  416     &       414     & 15      & e*   &       K2 Ve     &       2       &       4900    &       2.52    &       K2      &       4900     &  4900  \\
 6 &  TYC 6781-0415-1  &  409     &       426     & 13      & e*   &       G9 IVe    &       2       &       5410    &       4.69    &       G9.5    &       5330     &  5410  \\
 7 &  TYC 6803-0897-1  &  408     &       413     & 14      & a    &       \nodata   &      \nodata  &      \nodata  &       3.62    &       K0.5    &       5165     &  5165  \\
 8 &  TYC 6214-2384-1  &  397     &       398     & 14      & a    &       K1 IV     &       2       &       5080    &       2.48    &       K2      &       4900     &  5080  \\
 9 &  TYC 6806-0888-1  &  320     &       345     & 12      & a    &       G8 IV     &       2       &       5520    &       11.3    &       G3      &       5830     &  5520  \\
10 &  BD $+$04 3405B   &  233     &       242     & 5       & a    &     \nodata     &     \nodata   &      \nodata  &    $\ddagger$ &     \nodata   &       \nodata  &  6600$^{\star}$ \\
11 &  HD 144713        &  164     &       193     & 5       & a    &       F4        &       5       &       6590    &    $\ddagger$ &     \nodata   &       \nodata  &  6590  \\
12 &  HD 153439        &  180     &       198     & 5       & a    &       F5 V      &       3       &       6440    &    $\ddagger$ &     \nodata   &       \nodata  &  6440  \\ \hline

13 &  HD 148396        &  197     &       216     & 14      & a    &       K1/2 + F  &       3       &       5080    &       8.09    &       G8.5    &       5465     &  5080  \\                                                                                               
14 &  TYC 7334-0429-1  &  368     &       378     & 15      & a    &       K2e       &       2       &       4900    &       3.99    &       K0      &       5250     &  4900  \\
15 &  TYC 6817-1757-1  &  274     &       244     & 13      & e*   &       K0 Ve     &       2       &       5250    &       6.03    &       G9      &       5410     &  5250  \\
16 &  HD 157310        &  52      &       76      & 1       & a    &    A7 II/III    &       5       &       7850    &    $\ddagger$ &     \nodata   &       \nodata  &  7850  \\
17 &  HD 142016        &  20      &       57      & 0       & a    &     A4 IV/V     &       3       &       8460    &    $\ddagger$ &     \nodata   &       \nodata  &  8460  \\
18 &  CD -25 11942     &  307     &       324     & 13      & a    &       K0 IV     &       2       &       5250    &       8.72    &       G8.5    &       \nodata  &  5250  \\ 
19 &  2MASS J17215666-2010498 &  223     &       226     & 20      & e*   &       \nodata   &     \nodata   &       \nodata &       0.58    &       K7.5    &       4060     &  4060  \\ 
20 &  TYC 6790-1227-1  &  324     &       344     & 13      & a    &       G9 IV     &       2       &       5410    &       4.53    &       G9.5    &       5330     &  5410  \\
21 &  TYC 7346-1182-1  &  256     &       262     & 12      & a    &       G8 V      &       2       &       5520    &       5.94    &       G9      &       5410     &  5520  \\ \hline
                                                                                               
   &  BD -06 4414      &  \nodata &       \nodata & 10      & c    &       G5        &       1       &       5770    &    $\ddagger$ &       \nodata &       \nodata  & 5770   \\
   &  HD 137396        &  47      &       77      & 4       & a    &    F2/3 IV/V    &       5       &  6890 / 6740  &    $\ddagger$ &       \nodata &       \nodata  & 6815   \\
   &  HD 138969        &  113     &       109     & 9       & c    &       G1 V      &       4       &       5945    &       4.68    &       K9.5    &       3955     & 5945   \\
   &  HD 140574        &  \nodata &       \nodata & 1       & a    &    A9 V / A3    &     5 / 1     &  7390 / 8720  &    $\ddagger$ &       \nodata &       \nodata  & 8055   \\
   &  HD 141612        &  128     &       120     & 10      & a    &       G5 V      &       5       &       5770    &       4.25    &       K0      &       5250     & 5770   \\
   &  HD 141693        &  \nodata &       \nodata & \nodata & a    &       A0        &       1       &       9520    &    $\ddagger$ &       \nodata &       \nodata  & 9520   \\
   &  HD 141813        &  201     &       187     & 14      & a    & G8/K2 III + F/G &       3       &  5520 / 4900  &       31.4    &       $>$G2   &       $>$5860  & 5210   \\
   &  HD 142987        &  150     &       216     & 10      & e*   &  G3/6 /  G5     &    4  / 6     &5830-5700/5770 &    $\ddagger$ &     \nodata   &       \nodata  & 5770   \\
   &  HD 143332        &  \nodata &       \nodata & 5       & a    &      F5 V       &       5       &       6440    &       4.01    &       K0      &       5250     & 6440   \\
   &  HD 143358        &  196     &       215     & 9       & a    &     G1 / G2V    &       3       &  5945 / 5860  &    $\ddagger$ &       \nodata &       \nodata  & 5900   \\ 
   &  HD 143810        &  66      &       67      & 6       & a    &     F5/6 V      &       5       &  6440 / 6360  &       9.25    &       G8      &       5520     & 6400   \\
   &  HD 144393        &  \nodata &       \nodata & 7       & a    &     F7/8 V      &       4       &  6280 / 6200  &       13.8    &       $>$G2   &       $>$5860  & 6240   \\
   &  HD 144726        &  29      &       36      & 5       & a    &      F5 V       &       5       &       6440    &    $\ddagger$ &       \nodata &       \nodata  & 6440   \\
   &  HD 144732        &  178     &       195     & 9       & a    & G0 V / G0       &     3  / 6    &       6030    &       27.1    &       $>$G2   &       $>$5860  & 6030   \\
   &  HD 145169        &  51      &       46      & 10      & a    &       G3 V      &       5       &       5830    &       13.5    &       $>$G2   &       $>$5945  & 5830   \\
   &  HD 145551        &  26      &       25      & 6       & a    &       F5/6 V    &       5       &  6440 / 6360  &       9.17    &       G8.5    &       5465     & 6400   \\
   &  HD 148982        &  181     &       216     & 8       & a    &       F8 / G0   &       3       &  6200 / 6030  &    $\ddagger$ &       \nodata &       \nodata  & 6115   \\
   &  HD 154922        &  \nodata & 12$^{\dagger}$ & 1       & a    &       A7 III    &       4       &       7850    &    $\ddagger$ &       \nodata &       \nodata  & 7850   \\
   &  HIP 75685        &  99      &       99      & 11      & a    &       G6        &       1       &       5700    &       9.05    &       G8.5    &       5465     & 5700   \\
   &  HIP 79354        &  38      &       41      & 8       & a    &       F8        &       1       &       6200    &       2.87    &       K1      &       5080     & 6200   \\
   &  RHS 48           &  157     &       153     & 15      & e*   &       K2        &       1       &       4900    &       1.77    &       K2      &       4900     & 4900   \\
   &  TYC 0376-0769-1  &  12      &       21      & \nodata & a    &     \nodata     &    \nodata    &      \nodata  &    $\ddagger$ &       \nodata &       \nodata  & \nodata \\
   &  TYC 0909-0125-1  &  53      &       54      & 14      & a    &     \nodata     &     \nodata   &      \nodata  &       4.19    &       K0.5    &       5165     & 5165   \\
   &  TYC 0937-0754-1  &  99      &       97      & 16      & c    &     \nodata     &     \nodata   &      \nodata  &       1.89    &       K2.5    &       4815     & 4815   \\
   &  TYC 0976-1617-1  &  \nodata &       \nodata & 3       & a    &       A9 V      &       5       &       7390    &    $\ddagger$ &       \nodata &       \nodata  & 7390   \\
   &  TYC 5003-0138-1  &  79      &       79      & 16      & e*,p &     \nodata     &     \nodata   &      \nodata  &       1.85    &       K2.5    &       4815     & 4815   \\
   &  TYC 5022-0263-1  &  \nodata &       \nodata & \nodata & o    &    \nodata      &    \nodata    &     \nodata   &    $\ddagger$ &       \nodata &       \nodata  & \nodata \\
   &  TYC 5668-0365-1  &  61      &       70      & \nodata & e    &     \nodata     &     \nodata   &      \nodata  &    $\ddagger$ &       \nodata &       \nodata  & \nodata \\
   &  TYC 6141-0525-1  &  \nodata &       \nodata & \nodata & a    &    \nodata      &    \nodata    &     \nodata   &    $\ddagger$ &       \nodata &       \nodata  & \nodata \\
   &  TYC 6215-0184-1  &  49      &       52      & 15      & e*   &       K2IVe     &       2       &       4900    &    $\ddagger$ &       \nodata &       \nodata  & 4900   \\
\enddata
\tablecomments{We report here two Li $\lambda$6707 measurements- ``Integ.'' in column 3 refers to direct integration over the 
line profile, and ``Gfit'' in column 4 indicates the result of fitting a Gaussian to the absorption feature.  
In column 5, we also report contamination (denoted ``Ctmn.'') of the Li {\sc I} line; see \S\ \ref{lithium} 
for description of its derivation.  $\dagger$ Blended line; result indicates Gaussian feature fit to Li {\sc I} 
in deblending.  $\ddagger$ denotes cases in which the line ratio could not be measured from the spectrum either 
due to extreme rotational broadening or the lack of presence of either or both lines in question.
$^{\star}$ Effective temperature determined via interpolation of dereddened $H-K$ color over the 
color-effective temperature relationship of \citet{Kenyon:1995}, see Fig. \ref{fig_tempscale}.
}
\tablenotetext{1.}{Indicator flags are defined as follows: Absorption, a; core filling observed, 
c; double peaked emission, e*; P-Cygni like feature, p; emission with overlaid absorption, o.}
\tablenotetext{2.}{Spectral types drawn from the following sources:
typed by A.J. Weinberger using KAST low-resolution spectrograph, 1; 
\citet[SACY,][]{Torres:2006}, 2; Michigan spectral atlas \citep{Houk:1982,Houk:1988,Houk:1999}, 
3, 4, and 5, respectively; HD Catalog spectral type, 6.}
\end{deluxetable}
\end{landscape}

\begin{landscape}
\begin{deluxetable}{llcc|cccc|cccc}
\tabletypesize{\scriptsize}
\tablecaption{Kinematic Analysis of Membership Probability\label{Astrometry}}
\tablewidth{0pt}
\tablehead{\colhead{Plot}& \colhead{Object Name} & \colhead{$V sin i$}     & \colhead{Measured v$_{r}$} & \colhead{Predicted v$_{r}$} & \colhead{$\mu_{\tau}$}      & \colhead{Comovement}  & \colhead{Distance} & \colhead{Predicted v$_{r}$} & \colhead{$\mu_{\tau}$}      & \colhead{Comovement}  & \colhead{Distance} \\
           \colhead{ID}  & \colhead{}            & \colhead{[km s$^{-1}$]} & \colhead{[km s$^{-1}$]}     & \colhead{[km s$^{-1}$]}    & \colhead{[mas yr$^{-1}$]}    & \colhead{Probability} & \colhead{[pc]}     & \colhead{[km s$^{-1}$]}     & \colhead{[mas yr$^{-1}$]}  & \colhead{Probability} & \colhead{[pc]} }
\startdata
    &                	      &               &                       &    \multicolumn{4}{l}{Upper Sco Velocity Model}                                  &\multicolumn{4}{l}{UCL Velocity Model}                                               \\ \hline
 A  & HD 141569  	      &	  \nodata     &	      -6$\pm$5	      &    -9.4      &     -4.9 $\pm$   1.1     &     15.5    &    127 $\pm$   11        &    -8.1      &  -0.7 $\pm$   1.1         &          95.4     &     148 $\pm$  11    \\
 1  & TYC 6242-0104-1         &        14     &       -7.5$\pm$1.9    &    -7.5      &      -8.4$\pm$    3.6   &     9.0     &    228 $\pm$   59        &     -7.5     &   -5.1$\pm$    3.7        &           41.7    &     238 $\pm$  56    \\
 2  & TYC 6191-0552-1         &       $<$10   &        3.3$\pm$3.0    &      -6.1    &      -4.2$\pm$    3.4    &    56.4     &    151 $\pm$   25        &    -4.7      &    0.0 $\pm$   3.4        & 	     100.0    &     171 $\pm$  26    \\
 3  & TYC 6234-1287-1         &       22      &       -5.8$\pm$1.9    &      -8.7    &      -1.7$\pm$    2.6    &    92.0     &     87 $\pm$    9        &    -8.8      &    6.5 $\pm$   2.6        & 	      22.7    &     101 $\pm$   9    \\
 4  & TYC 7312-0236-1         &       24      &        4.7$\pm$1.7    &      -0.7    &      -6.3$\pm$    3.3    &    26.2     &    140 $\pm$   19        &     1.4      &   -1.8 $\pm$   3.3        & 	      88.7    &     152 $\pm$  20    \\
 5  & TYC 7327-0689-1         &       16      &       -0.7$\pm$1.7    &      -1.8    &      -4.0$\pm$    3.2    &    61.3     &    118 $\pm$   14        &     0.0      &    1.6 $\pm$   3.2        & 	      91.8    &     130 $\pm$  15    \\
 6  & TYC 6781-0415-1         &       29      &       -3.0$\pm$2.1    &      -3.4    &      -2.6$\pm$    2.6    &    77.8     &    109 $\pm$   11        &    -1.7      &    3.1 $\pm$   2.6        & 	      65.8    &     123 $\pm$  12    \\
 7  & TYC 6803-0897-1         &       22      &       -4.2$\pm$1.9    &      -4.3    &      -4.3$\pm$    2.5    &    46.8     &    120 $\pm$   13        &    -3.3      &    1.5 $\pm$   2.5        & 	      89.3    &     134 $\pm$  13    \\
 8  & TYC 6214-2384-1         &       15      &       -3.3$\pm$1.4    &      -5.5    &      -6.8$\pm$    3.5    &    27.8     &    121 $\pm$   17        &    -4.4      &   -1.2 $\pm$   3.6        & 	      95.6    &     134 $\pm$  18    \\
 9  & TYC 6806-0888-1         &       43      &       -2.2$\pm$1.8    &      -3.7    &      -2.2$\pm$    3.0    &    84.6     &    127 $\pm$   15        &    -2.7      &    3.3 $\pm$   2.9        & 	      64.6    &     144 $\pm$  16    \\
10  & BD $+$04 3405B$^{\ddagger}$  &       39      &       -14.1$\pm$2.2   &     -13.9    &      -0.6$\pm$    1.5    &    96.2     &    169 $\pm$   23        &   -14.3      &    3.3 $\pm$   1.6        & 	      29.2    &     207 $\pm$  27    \\
11  & HD 144713               &       $>$70   &       0.6$\pm$4.4     &      -4.4    &      -1.9$\pm$    1.7    &    72.7     &    163 $\pm$   16        &    -3.1      &    2.1 $\pm$   1.7        & 	      64.0    &     184 $\pm$  17    \\
12  & HD 153439               &       56      &       -4.3$\pm$1.6    &      -5.0    &       1.4$\pm$    1.6    &    88.3     &    130 $\pm$   12        &    -4.5      &    6.8 $\pm$   1.6        & 	       2.0    &     151 $\pm$  13    \\ \hline

13  & HD 148396               &       36      &       -0.2$\pm$2.1    &      -3.4    &      -0.9$\pm$    2.4    &    94.4     &    231 $\pm$   37        &    -2.4      &    2.1 $\pm$   2.4        & 	      72.6    &     261 $\pm$  41    \\			                               				  	 		           	     	     	            		    	            			          	      			     	     		          		  			   	       		       			     				   					       		  
14  & TYC 7334-0429-1         &       27      &       -4.3$\pm$1.5    &      -2.2    &      -5.2$\pm$    2.2    &    25.3     &    129 $\pm$   13        &    -0.8      &    0.1 $\pm$   2.2        & 	      99.9    &     142 $\pm$  14    \\
15  & TYC 6817-1757-1         &       11      &       8.4$\pm$1.9     &      -4.8    &      -7.4$\pm$    2.8    &     3.8     &    379 $\pm$   99        &    -4.1      &   -5.4 $\pm$   2.7        & 	      16.5    &     382 $\pm$  90    \\
16  & HD 157310               &       $>$70   &-22.4$\pm$2.3$\dagger$ &     -13.9    &      -1.6$\pm$    1.5    &    74.1     &    204 $\pm$   28        &   -14.3      &    1.7 $\pm$   1.5        & 	      67.7    &     245 $\pm$  32    \\
17  & HD 142016               &       \nodata &-17.8$\pm$2.3$\dagger$ &      -2.3    &      -6.2$\pm$    1.2    &    22.6     &     85 $\pm$    7        &    -0.7      &    1.7 $\pm$   1.2        & 	      87.3    &      94 $\pm$   6    \\
18  & CD-25 11942             &       53      &       -6.5$\pm$1.5    &      -5.7    &      -2.1$\pm$    2.0    &    78.0     &    127 $\pm$   12        &    -5.4      &    3.6 $\pm$   2.0        & 	      43.4    &     144 $\pm$  13    \\
19  & 2MASS J17215666-2010498  &       $<$10   &       -7.3$\pm$1.9    &      -7.5    &      -8.4$\pm$    3.6    &     9.0     &    228 $\pm$   59        &    -7.5      &   -5.1 $\pm$   3.7        & 	      41.7    &     238 $\pm$  56    \\
20  & TYC 6790-1227-1         &       33      &       2.0$\pm$2.0     &      -2.6    &      -6.1$\pm$    2.7    &    23.7     &    121 $\pm$   13        &    -1.0      &   -0.7 $\pm$   2.6        & 	      98.1    &     133 $\pm$  13    \\
21  & TYC 7346-1182-1         &       28      &       3.7$\pm$1.7     &      -3.6    &      -4.8$\pm$    2.3    &    31.8     &    129 $\pm$   13        &    -2.8      &    0.7 $\pm$   2.3        & 	      97.1    &     143 $\pm$  14    \\ \hline

\enddata
\tablecomments{$\dagger$Very broadened, featureless spectra.  Radial velocity derived 
via centroid measurement of the H$\alpha$ line.  Uncertainty reflects an 
assumed centroid measurement error of 0.05\AA.  For a full discussion of error 
analysis, see $\S$ \ref{rv}.}
\end{deluxetable}
\end{landscape}

\begin{deluxetable}{llcccc}
\tabletypesize{\scriptsize}
\tablecaption{Association Notes\label{Membership}}
\tablewidth{0pt}
\tablehead{\colhead{Plot}  &  \colhead{Object} & \colhead{Spatial} & \colhead{Velocity Model / } & \colhead{US/UCL V$_{r}$} &  \colhead{Determined}\\
           \colhead{ID}    &  \colhead{Name}   & \colhead{Match}   & \colhead{Probability}       & \colhead{Prediction}     &  \colhead{Membership}  }
\startdata
 1 &  TYC 6242-0104-1         & ?      &  US / 9\%     & 1$\sigma$/1$\sigma$  & I       \\
 2 &  TYC 6191-0552           & US     & US / 56\%     & no/3$\sigma$         & US      \\
 3 &  TYC 6234-1287-1         & ?      & US / 92\%     & 2$\sigma$/2$\sigma$  & US?     \\
 4 &  TYC 7312-0236-1         & UCL    & UCL / 89\%    & no/2$\sigma$         & UCL     \\
 5 &  TYC 7327-0689-1         & UCL    & UCL / 92\%    & 1$\sigma$/1$\sigma$  & UCL     \\
 6 &  TYC 6781-0415-1         & US     & US / 78\%     & 1$\sigma$/1$\sigma$  & US      \\
 7 &  TYC 6803-0897-1         & US     & US / 47\%     & 1$\sigma$/1$\sigma$  & US      \\
 8 &  TYC 6214-2384-1         & US     & US / 28\%     & 2$\sigma$/1$\sigma$  & US      \\
 9 &  TYC 6806-0888-1         & US     & US / 85\%     & 1$\sigma$/1$\sigma$  & US      \\
10 &  BD $+$04 3405B          & ?      & US / 96\%     & 1$\sigma$/1$\sigma$  & I       \\
11 &  HD 144713               & US     & US / 73\%     & 2$\sigma$/1$\sigma$  & US      \\
12 &  HD 153439               & ?      & US / 88\%     & 1$\sigma$/1$\sigma$  & US?     \\ \hline

13 &  HD 148396               & US     & US / 94\%     & 2$\sigma$/2$\sigma$  & US      \\
14 &  TYC 7334-0429-1         & UCL    & UCL / 100\%   & 2$\sigma$/3$\sigma$  & UCL     \\
15 &  TYC 6817-1757-1         & US     &  US / 4\%     & no/no                & I       \\
16 &  HD 157310               & ?      & US / 78\%     & no/no                & I       \\
17 &  HD 142016               & UCL    & UCL / 87\%    & no/no                & UCL     \\
18 &  CD-25 11942             & ?      & US / 78\%     & 2$\sigma$/2$\sigma$  & US?     \\
19 &  2MASS J17215666-2010498  & ?      &  US / 9\%     & 1$\sigma$/1$\sigma$  & I       \\
20 &  TYC 6790-1227-1         & UCL    & UCL / 98\%    & 3$\sigma$/2$\sigma$  & UCL     \\
21 &  TYC 7346-1182-1         & US     & US / 32\%     & no/no                & US      \\
\enddata
\tablecomments{In column three, we note with which moving group each object is spatially 
consistent.  Column four summarizes the results of velocity vector modeling and shows the 
vector with which each object had the highest comovement probability.  In column five, we 
report the models which predicted radial velocities within 2$\sigma$ of our 
measured values, and in the rightmost column we comment on membership; I denotes  
``indeterminate''.  See text ($\S$ \ref{summconc}) for further discussion of our final 
membership determinations.}
\end{deluxetable}

\clearpage

\bibliographystyle{apj}
\bibliography{refs} 

\end{document}